# Z-Fold Wing Aeroelasticity: Compositional Modeling, 1:2 Double-Hopf Dynamics, and Nonlinear Stiffness Effects

Gelin Chen*, Haolin Yang†, Chen Song‡, and Chao Yang§

*School of Aeronautics, Beihang University, Beijing 100191, People's Republic of China*

**Folding changes the linearized aeroelastic spectrum and can switch which mode becomes unstable first, with consequences for local postflutter interactions. This study formulates a three-component Z-fold wing by assigning every aerodynamic station to a structural component and a material coordinate. The same attachment map generates surface motion and returns pressure loads through virtual work. Geometrically exact component dynamics, an explicit-wake unsteady vortex-lattice model, and block-structured descriptor assembly preserve the physical paths of configuration actions. A two-parameter flutter analysis shows that a smooth flutter-speed envelope conceals a high–low–high sequence of controlling neutral branches, expressed as a critical-frequency valley and a redistribution of component deformation. Numerical continuation locates a near-1:2 double-Hopf point. Within a local model retaining quadratic and cubic structural restoring forces with aerodynamic and inertial operators fixed at the scheduling point, the cubic normal form captures selected 26-state observations and admits high-frequency-dominant and mixed phase-locked periodic solutions. The formulation links configuration-dependent flutter-mode identity to local resonant dynamics and supports blockwise sensitivity and design reasoning.**

## Nomenclature

| Symbol | Definition |
|---|---|
| $B_i, H_i(s,t)$ | root pose and material-section pose of component $i$ in $SE(3)$ |
| $E, L$ | descriptor and linear state matrices, $E\dot{x} = Lx$ |

* Ph.D. Candidate, School of Aeronautics, Beihang University; sy2405230@buaa.edu.cn.

† Ph.D. Candidate, School of Aeronautic Science and Engineering; yanghaolin@buaa.edu.cn.

‡ Associate Professor, School of Aeronautics, Beihang University; songchen@buaa.edu.cn.

§ Professor, School of Aeronautics, Beihang University; yangchao@buaa.edu.cn.

| Symbol | Definition |
| --- | --- |
| $\mathcal{P}$ | configuration-dependent primitive blocks used to assemble (E,L) |
| $A_w, B_w, C_w, D_w, \dot{D}_w$ | wake, input, output, direct, and pressure-rate operators |
| $M_s, C_s, K_s$ | structural mass, damping, and tangent-stiffness matrices |
| $N_q, N_v$ | displacement and velocity maps into normal boundary velocity |
| $U, V^*$ | dimensional speed and $U/(b\omega_0)$, with $b = c/2$ |
| $v_j, w_j$ | right and left generalized eigenvectors |
| $X_S, X_A, r_{SA}$ | structural station, aerodynamic station, and $X_A - X_S$ |
| $d_e, \sigma_e$ | element relative logarithm and endpoint-preserving internal coordinate |
| $p_{j,g}$ | left–right participation assigned to state block $g$ |
| $q_{B_i}, \theta_{12}, \theta_{23}$ | retained component and fold coordinates |
| $u, v, n, \Delta p^{\text{phys}}$ | chordwise/spanwise surface coordinates, oriented normal, and $p_{\text{lower}} - p_{\text{upper}}$ |
| $x = [q^T, v^T, z_w^T]^T$ | 26-state structural–wake vector, $q, v \in \mathbb{R}^{11}$, $z_w \in \mathbb{R}^4$ |
| $z_L, z_H$ | complex center coordinates normalized by $\rho_{\text{work}}$ |
| $\Gamma_b, \Gamma_{w0}, \Gamma_w$ | bound, newly shed, and transported-wake circulations |
| $\beta_L, \beta_H, \delta$ | unfolded growth rates and detuning $\omega_H - 2\omega_L$ |
| $\epsilon_A$ | nondimensional chordwise structural-axis offset |
| $\lambda_{L,H} = \alpha_{L,H} + i\omega_{L,H}$ | low- and high-frequency critical eigenvalues |
| $\phi$ | symmetric reference fold angle |
| $r_L, r_H, \psi$ | modal amplitudes and resonant phase, $\psi = \arg z_H - 2\arg z_L$ |
| $\xi_i, \eta_i$ | body strain and material velocity |
| $\rho_F$ | Floquet multiplier |

## I. Introduction

Folding changes planform geometry, structural load paths, local surface-normal velocity, and the transfer of aerodynamic forces into the primary structure. These changes modify the coupled linearized operator and can switch which aeroelastic mode becomes unstable first. The critical frequency and modal deformation pattern may therefore change even when the minimum flutter-speed curve remains visually smooth. Morphing-aircraft reviews and folding-wing studies place this configuration-dependent stability problem among the central engineering issues of adaptive lifting surfaces [1, 2].

The structural–aerodynamic relation is familiar in conventional wing idealizations: an aerodynamic panel follows a section supported by ribs and spars, and its pressure load returns through that structural load path. A component-attached formulation makes this implicit relation explicit. Each aerodynamic station is assigned to one parent component and one material coordinate; the same attachment map generates station motion and returns pressure loads to structural coordinates through virtual work. The representation can also accommodate an in-plane movable trailing-edge segment by transporting its parent pose and can describe a variable-camber surface by allowing the attachment field to depend on a camber coordinate. These are representation routes rather than additional demonstrations in the present Z-fold example; related camber-morphing structures and aeroelastic models illustrate the broader class of structural–surface relations [3, 4].

The folding-wing literature provides the immediate engineering baseline. Theoretical–experimental work related fold geometry and connection properties to vibration and flutter [5]; continuum and nonlinear three-dimensional formulations introduced parameterized structural and vortex-lattice descriptions [6, 7]; and multisegment models connected folding-wing equations to measured flutter speeds and frequencies [8]. Partitioned finite-element and unsteady-vortex-lattice simulations treated nonlinear response across flight configurations [9]. Xie et al. coupled a continuous-time state-space aerodynamic model to a three-segment Z-shaped wing using the component velocity projected onto the local aerodynamic-surface normal [10]. A recent rigid–elastic aeroservoelastic model with fold angle and Mach number as scheduling variables likewise reported configuration-dependent flutter degradation and mode switching [11].

Mode switching is not completely described by a scalar flutter boundary. Parameterized folding-wing calculations have shown changes of the controlling mode, critical shape, and reported flutter frequency [12]; flexible-multibody studies have addressed transient response during folding [13]; and cubic or bilinear hinge laws have been linked to limit cycles and parameter-dependent nonlinear response [14, 15]. Nonlinear analyses of Z-shaped wings include computational and experimental studies and a 1:1 internal-resonance model [16, 17]. If branch-specific neutral speeds are smooth, their minimum can still change controlling branch and boundary tangent. Branch identity, left/right eigenvectors, and first-order spectral information are therefore relevant to stability-margin and local design reasoning, rather than being only numerical labels attached to a parameter sweep.

The required aerodynamic and reduction tools are established. The unsteady vortex-lattice method represents three-dimensional potential-flow loading with an explicit convected wake [18], and state-space realizations separate dynamic-wake feedback, instantaneous loading, and pressure-rate contributions for arbitrary kinematics [19]. Local models have been represented across morphing operating regions [20], and parameterized wake reduction lowers the computational cost of retaining explicit wake states [21]. Generalized-eigenvalue perturbation then relates block derivatives and left/right eigenvectors to branch-specific eigenvalue actions [22].

For the local nonlinear dynamics, resonant double-Hopf bifurcations organize the interaction of two oscillatory pairs approaching neutrality. Aeroelastic examples include a tuned-mass-damper oscillator [23], a normal-form representation of the Goland wing [24], and a Hopf–Hopf analysis of a three-degree-of-freedom airfoil [25]. General 1:2 and 1:3 resonant-Hopf analyses identify the quadratic transfer terms and quotient phase dynamics activated near commensurate frequencies [26]. Taken together, these studies motivate treating resonant double-Hopf dynamics as the local nonlinear extension of a branch-resolved, configuration-dependent flutter problem. The same configuration changes that switch which aeroelastic branch becomes unstable first can also alter the critical left and right eigenspaces through which the retained quadratic and cubic structural restoring-force terms enter the resonant normal form.

This paper develops three connected contributions for a symmetric three-component Z-fold wing. First, a component-attached compositional model exposes how fold angle and structural-axis offset act through kinematic, aerodynamic, structural, and load-transfer blocks. Second, a branch-resolved flutter analysis identifies a high–low–high change of the controlling neutral branch beneath the smooth flutter envelope and relates it to component-level modal redistribution. Third, numerical continuation and a local center-manifold reduction connect the competing branches to a near-1:2 double-Hopf point and to the response classes admitted by the cubic normal form. Spatial and reduced-order convergence assessments support the baseline analysis model, while the appendices provide the derivations needed to close the mathematical route.

## II. Compositional Z-Fold Aeroelastic Formulation

The Z-fold is represented as a directed serial chain with component nodes $B_1, B_2, B_3$ and revolute-joint edges 12 and 23. Each nonroot component has one parent: $B_1$ is the root, $B_2$ is attached to $B_1$, and $B_3$ is attached to $B_2$. Twists and virtual displacements are transported from parent to child, whereas force and moment one-forms are pulled back from child to parent. This work-conjugate organization follows geometrically exact beam and flexible-multibody mechanics [27–31]. An aerodynamic material chart is assigned to each component: $u$ runs from leading edge to trailing edge, $v$ runs from component root to tip, and the ordered tangents define the oriented surface normal

$$n = \frac{X_{,u} \times X_{,v}}{\parallel X_{,u} \times X_{,v} \parallel} \qquad (1)$$

The reference configuration and the two discretizations are shown in Fig. 1. The second panel distinguishes beam stations, aerodynamic panels, control points, and selected normals defined by Eq. (1).

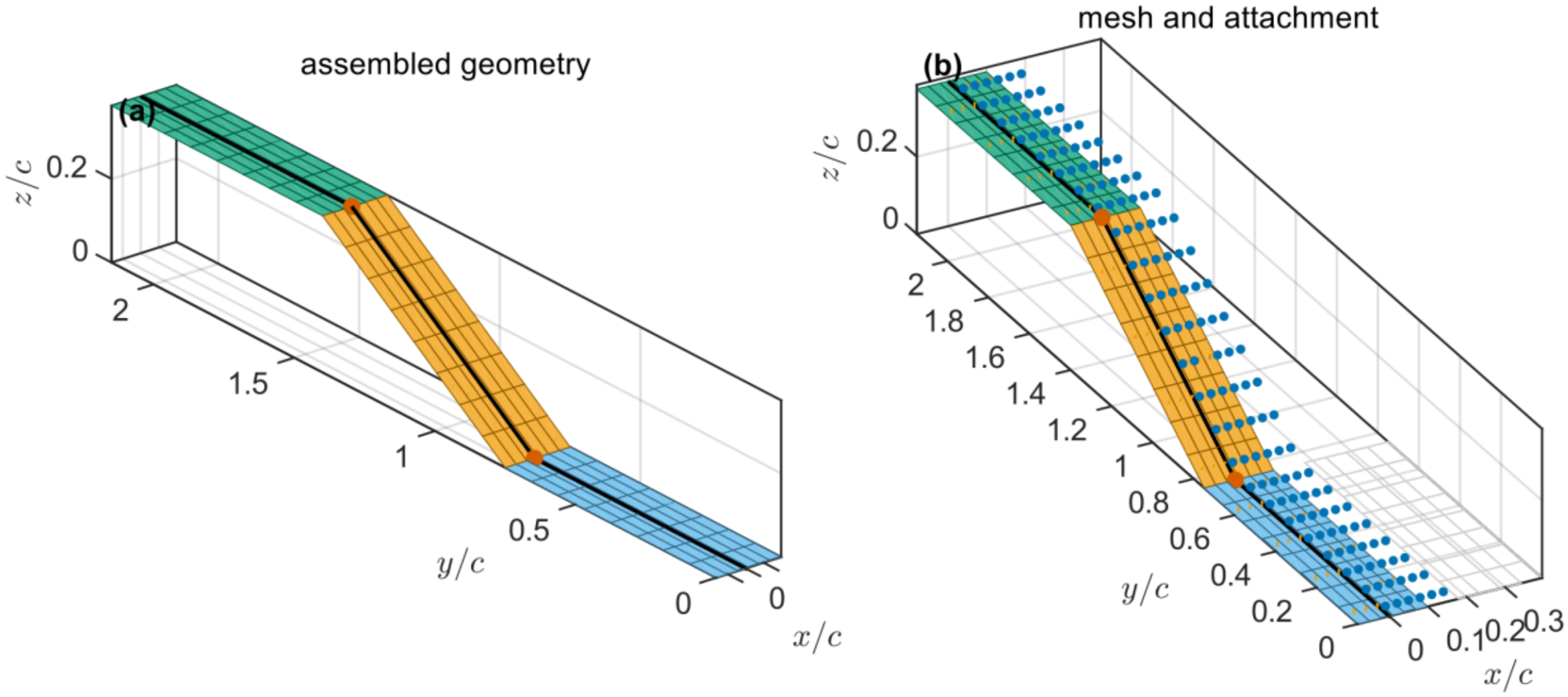

**Fig. 1. Z-fold geometry and attachment: a) component surfaces; b) beam stations, control points, and normals. Along increasing $y/c$, components are $B_1, B_2, B_3$.**

**A. Geometry, coordinates, and discretization**

The material-section pose of component $i$ is factored into root transport, reference geometry, and elastic deformation,

$$H_i(s,t) = B_i(t) g_i(s) E_i(s,t), \qquad s \in [0, L_i], \qquad H_i \in SE(3) \tag{2}$$

Here $B_i$ carries the motion inherited from upstream components, $g_i$ fixes the undeformed component geometry, and $E_i = \exp(\hat{\sigma}_i)$ contains elastic motion relative to that geometry. The left-trivialized derivative along the material coordinate defines the body strain, and the left-trivialized time derivative defines the material velocity:

$$\xi_i = \left(H_i^{-1} H_{i,s}\right)^{\vee}, \qquad \eta_i = \left(H_i^{-1} H_{i,t}\right)^{\vee} \tag{3}$$

Rigid transport by $B_i$ contributes to material velocity, inertia, and external work while leaving the component strain invariant. For a beam element with endpoint poses $H_a, H_b$, the exact endpoint relative pose is retained through the logarithm

$$d_e = \log(H_a^{-1} H_b)^{\vee} \tag{4}$$

which determines the element strain and elastic potential. The internal coordinate $\sigma_e(s)$ reconstructs the same endpoint relative pose and uses the selected strain-interpolation order. Only perturbations about the finite reference pose are expanded. Appendix A gives the endpoint-preserving expansion and its order consistency.

### B. Geometrically exact component dynamics

For a variation $\delta H_i = H_i \hat{\zeta}_i$, the exact kinematic variations are

$$\delta\eta_i = \dot{\zeta}_i + \mathrm{ad}_{\eta_i}\zeta_i, \qquad \delta\xi_i = \zeta_{i,s} + \mathrm{ad}_{\xi_i}\zeta_i \tag{5}$$

The elastic potential of a component is

$$\mathcal{U}_i = \frac{1}{2}\int_0^{L_i} (\xi_i - \xi_i^0)^T\, \mathbb{K}_i(s)(\xi_i - \xi_i^0)\, ds \tag{6}$$

With body momentum $\mu_i = \partial\mathcal{T}_i/\,\partial\eta_i$, internal-force one-form $n_i = \partial\mathcal{U}_i/\,\partial\xi_i$, and distributed external one-form $f_i$, the component variation is

$$\delta S_i = \iint \left[\langle \mu_i, \dot{\zeta}_i + \mathrm{ad}_{\eta_i}\zeta_i\rangle - \langle n_i, \zeta_{i,s} + \mathrm{ad}_{\xi_i}\zeta_i\rangle + \langle f_i, \zeta_i\rangle\right] ds\, dt \tag{7}$$

Integration by parts gives the Euler–Poincaré balance [27, 31],

$$\mathcal{R}_i = -\dot{\mu}_i + \mathrm{ad}^*_{\eta_i}\mu_i + n_{i,s} - \mathrm{ad}^*_{\xi_i} n_i + f_i = 0 \tag{8}$$

The material boundary term is a wrench one-form. It supplies the port quantity required for assembling the component chain and makes the same weak form valid for an isolated component and for a component embedded in the Z-fold.

### C. Component transmission and cotangent pullback

At a component port, the internally reconstructed velocity is affine in the root twist and retained elastic rates,

$$V_p = Z_p(q)V_B + Y_p(q)\dot{q} \tag{9}$$

For an edge that joins this port to a child root,

$$V_{B_c} = H_e(q,\theta_e)V_p + J_e(q,\theta_e)\dot{\theta}_e, \qquad \zeta_{B_c} = H_e\zeta_p + J_e\delta\theta_e \tag{10}$$

If a child port carries the wrench one-form $\mathcal{B}_c$, equality of virtual work gives the cotangent pullback

$$\mathcal{B}_p = H_e^T\mathcal{B}_c, \qquad R_{\theta_e} = J_e^T\mathcal{B}_c, \qquad \mathcal{B}_c^T\zeta_{B_c} = \mathcal{B}_p^T\zeta_p + R_{\theta_e}\delta\theta_e \tag{11}$$

For the present chain, the component-3 wrench is first pulled through joint 23 and added to the component-2 port wrench. That sum is then pulled through joint 12 and added to the component-1 port wrench. Substitution of Eqs. (9)–(10) into the component weak forms, followed by collection of the independent root, elastic, and joint variations, gives the global structural residual. Both joint angles remain dynamic generalized coordinates in this residual.

Figure 2(a) summarizes the assembled component-motion and wrench-transfer chain.

### D. UVLM closure and aerodynamic virtual work

A component attachment map assigns each aerodynamic material coordinate $(u, v)$ to a beam coordinate $s_i$, an associated section coordinate $\eta_i$, and an offset $\rho_i$:

$$\mathcal{A}_i\colon (u, v) \mapsto \big(s_i(u, v), \eta_i(u, v), \rho_i(s_i, \eta_i)\big), \qquad X_A = \pi_x \left[ H_i(s_i, t) \begin{pmatrix} \rho_i \\ 1 \end{pmatrix} \right] \tag{12}$$

Differentiation of $\mathcal{A}_i$ supplies surface tangents, the normal in Eq. (1), the point velocity $v_A$, and the virtual-displacement Jacobian $B_{x,p}$. The resulting derivatives of aerodynamic generalized loads with respect to structural displacement and velocity are consistent with analytical UVLM load linearizations for geometrically exact beams [32]. The normal-flow input is

$$u_n(q, \dot{q}) = n(q)^T \left[U_\infty - v_A(q, \dot{q})\right], \qquad \delta u_n = N_q(U)\, \delta q + N_v\, \delta \dot{q}, \qquad N_q(U) = U N_{q1} \tag{13}$$

The displacement contribution to $\delta u_n$ contains both station motion and the change in freestream projection caused by rotation of the local normal. Because both follow from the attached material map, the motion transfer and aerodynamic virtual-work pullback remain kinematically and work-conjugately consistent.

Let $\Gamma_b$, $\Gamma_{w0}$, and $\Gamma_w$ denote bound circulation, the newly shed trailing-edge row, and the transported wake. Standard UVLM collocation, Kutta shedding, and wake transport [18, 19, 33] yield

$$K_1 \Gamma_b + K_2 \Gamma_{w0} + K_3 \Gamma_w = b_n, \qquad b_{n,p} = -n_p^T \big(U_p - V_p\big) \tag{14a}$$

$$K_4 \Gamma_b + K_5 \Gamma_{w0} = 0, \qquad \dot{\Gamma}_w = K_6 \Gamma_w + K_7 \Gamma_{w0} \tag{14b}$$

Solving the algebraic blocks eliminates the bound circulation $\Gamma_b$ and newly shed row $\Gamma_{w0}$, leaving the transported wake state $\Gamma_w$ and the prescribed surface kinematics as inputs to pressure recovery. The recovered pressure includes circulation-rate and tangential-transport terms. Positive pressure difference is defined as lower-surface minus upper-surface pressure, and positive panel force acts along the oriented normal in Eq. (1):

$$\Delta p_p^{phys} = p_{lower,p} - p_{upper,p} = \rho_\infty \Pi_p, \qquad F_p = A_p \Delta p_p^{phys} n_p \tag{15}$$

Aerodynamic loading is returned to structural coordinates by virtual work,

$$\delta W_a = \sum_p F_p^T \, \delta x_p = \delta y^T \sum_p B_{x,p}^T \, F_p, \qquad Q_a = \sum_p B_{x,p}^T \, F_p \tag{16}$$

If a force is transferred from an aerodynamic point $X_A$ to a structural station $X_S$, the moment satisfies

$$m_S = m_A + r_{SA} \times f_A, \qquad r_{SA} = X_A - X_S, \qquad \epsilon_A = \frac{r_{SA} \cdot e_c}{c} \tag{17}$$

Thus $\epsilon_A$ changes the pressure lever arm and therefore the generalized force and moment map. Figure 2(b) summarizes the aerodynamic chain from attached-surface motion to pressure work.

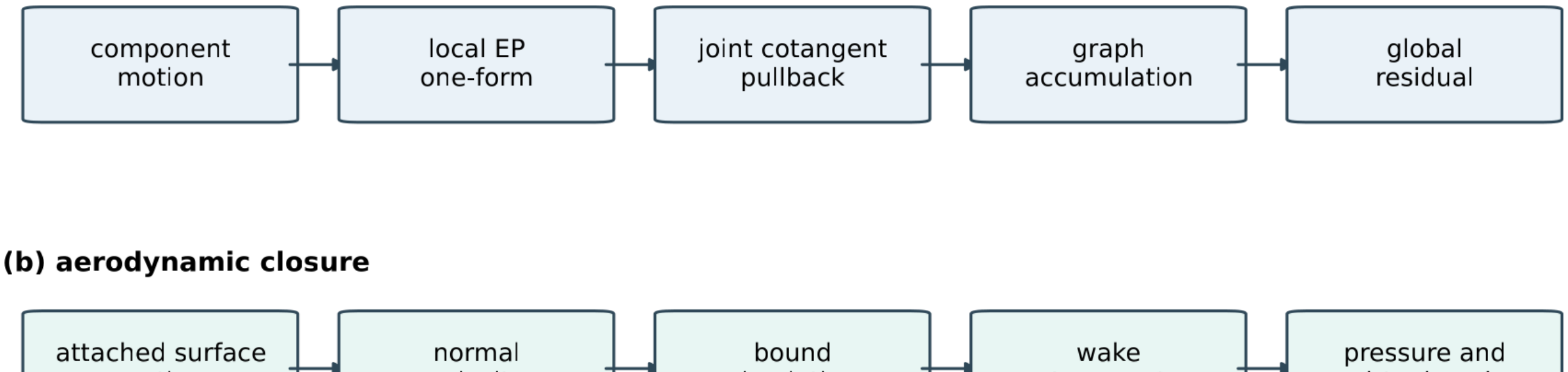

**Fig. 2. Compositional model closure: a) downstream kinematics and upstream cotangent accumulation; b) attached-surface motion, wake transport, pressure recovery, and virtual work.**

### E. Coupled descriptor system

Let $z_w$ denote a reduced wake coordinate. With the structural normal-velocity input from Eq. (13), an input–output realization is

$$\dot{z}_w = A_w z_w + B_w N_q q + B_w N_v \dot{q} \tag{18a}$$

$$Q_a = C_w z_w + D_w N_q q + D_w N_v \dot{q} + \dot{D}_w N_q \dot{q} + \dot{D}_w N_v \ddot{q} \tag{18b}$$

The terms in Eq. (18b) represent wake feedback, displacement- and velocity-induced loading, a pressure-rate contribution driven by displacement kinematics, and an acceleration term. The overdot in $\dot{D}_w$ labels the pressure-rate coefficient and is not a time derivative of $D_w$. For a fixed grid, convective time scales with speed, so the operators separate as

$$A_w = UA_1, \quad B_w = UB_1, \quad C_w = UC_1, \quad D_w = UD_1, \quad \dot{D}_w = \dot{D}_1 \tag{19}$$

Combining Eqs. (18a)–(18b) with $M_s\ddot{q} + C_s\dot{q} + K_s q = Q_a$ gives the configuration-consistent structural–aerodynamic linearization [34, 35],

$$M_e = M_s - \dot{D}_1 N_v, \quad C_e = C_s - U\big(D_1 N_v + \dot{D}_1 N_{q1}\big), \quad K_e = K_s - U^2 D_1 N_{q1} \tag{20}$$

For $x = [q^T, v^T, z_w^T]^T$, $v = \dot{q}$, the linear coupled system is

$$E\dot{x} = Lx, \quad E = \begin{bmatrix} I & 0 & 0 \\ 0 & M_e & 0 \\ 0 & 0 & I \end{bmatrix}, \quad L = \begin{bmatrix} 0 & I & 0 \\ -K_e & -C_e & UC_1 \\ U^2 B_1 N_{q1} & UB_1 N_v & UA_1 \end{bmatrix} \tag{21}$$

Wake reduction projects only the strictly proper dynamic subsystem. The direct block $D_1$, pressure-rate block $\dot{D}_1$, and their structural-input and aerodynamic-output coordinates are retained in their natural coordinates. This choice makes the reduction error an error in generalized aerodynamic force for a structural normal-velocity input [19, 21]. The right and left eigenvectors are normalized by

$$LV = EV\Lambda, \qquad W^H L = \Lambda W^H E, \qquad W^H E V = I \tag{22}$$

The same descriptor normalization is used for root tracking, component participation, and the center-manifold reduction. Residuals of Eq. (22) are included as algebraic verification metrics.

## III. Spatial and Reduced-Order Verification

Table 1 identifies the model used for each class of result. The baseline analysis grid is the fixed spatial discretization and retained-order pair used for the parameter-domain and nonlinear studies; it is distinct from the finest spatial reference used only for convergence assessment.

**Table 1. Model hierarchy for the reported results.**

| Result | Model |
|---|---|
| Spatial convergence | Finest common spatial reference with one discretization parameter varied in each sequence |
| Parameter-domain flutter and high–low–high sequence | Baseline analysis grid, $r_s = 9$, and $r_v = 4$ |
| Double-Hopf localization | Common-coordinate local primitive descriptor assembled as $(\mathcal{P}, U) \mapsto (E, L)$ |
| Time-domain reconstruction | Four-dimensional cubic normal form and 26-state reduced nonlinear model, identified separately |

| Periodic orbits and Floquet multipliers | Four-dimensional cubic normal form |
|---|---|

## A. Z-Fold specialization

The example contains three identical spanwise components connected in series by two fold joints. Each component supplies one in-plane bending family, one out-of-plane bending family, and one physical low-frequency torsional family to the retained structural space. With the two joint coordinates,

$$q = [q_{B1}^T, q_{B2}^T, q_{B3}^T, \theta_{12}, \theta_{23}]^T \in \mathbb{R}^{11} \quad (23)$$

The baseline structural basis contains nine elastic vectors, three associated with each component-local deformation family, and the baseline wake basis contains four states. The first-order dimension is therefore $11 + 11 + 4 = 26$. Table 2 lists the physical and numerical settings. The displayed parameter domain is $\epsilon_A \in [-0.35, -0.10]$ and $\phi \in [0, 60°]$.

**Table 2. Principal parameters and retained orders of the Z-fold example.**

| Quantity | Symbol | Value or policy |
|---|---|---|
| Flexible components | $n_c$ | 3 |
| Component length | $L$ | 0.75 m |
| Chord / half-chord | $c, b$ | 0.18 m / 0.09 m |
| Air density | $\rho_\infty$ | 1.225 kg m$^{-3}$ |
| Reference frequency | $f_0, \omega_0$ | 4.5 Hz, $2\pi f_0$ |
| Baseline beam / spanwise panel count | $N_e, N_v$ | 8, 8 per component |
| Chordwise panels | $N_u$ | 6 per component |
| Baseline wake rows / spacing / length | $N_w, \Delta x_w/c, L_w/c$ | 24 / 0.30 / 7.2 |
| Baseline structural / wake orders | $r_s, r_w$ | 9 / 4 |
| ROM reference orders | $(r_s, r_w)_{\mathrm{ref}}$ | (9,8) |
| First-order baseline dimension | $n_x$ | 26 |
| Modal damping ratio | $\zeta_s$ | 0.005 |
| Joint-frequency targets | $r_{J12}, r_{J23}$ | 1.05, 1.15 times $\omega_0$ |
| Parameter domain | $(\epsilon_A, \phi)$ | $[-0.35, -0.10] \times [0, 60°]$ |

The linear study retains the complete first-order attachment and normal maps, structural inertia, and explicit wake dynamics. The local nonlinear study uses the same aerodynamic linearization while retaining the structurally derived restoring force through cubic order.

## B. Shared spatial reference and one-dimensional refinement

Spatial refinement is evaluated at the representative point $(\epsilon_A, \phi) = (-0.20, 0°)$. A single common reference combines the largest structural and aligned spanwise grid, the longest wake, and the smallest wake spacing,

$$N_e = N_v = 16, \quad N_u = 6, \quad L_w/c = 9.6, \quad \Delta x_w/c = 0.15, \quad N_w = 64 \quad (24)$$

For either $y = V_f^*$ or $y = \omega_f/\omega_0$, the reported error is

$$e_y(h) = 100\frac{|y(h) - y_{ref}|}{|y_{ref}|}\ \%, \qquad y_{ref} = y(16{,}9.6{,}0.15) \tag{25}$$

Only one refinement parameter changes in each sequence. The other two discretization parameters are set equal to their reference values. During structural refinement, $N_v$ and the aerodynamic/wake spanwise stations follow $N_e$; this alignment makes the structural index the independent refinement parameter while preserving the attachment topology.

For the structural-grid study in Fig. 3, the analysis varies $N_e = N_v = 4,6,\ldots,16$ while retaining the longest and finest wake. Relative to Eq. (24), $N_e = N_v = 8$ gives $e_V = 1.0566\%$ and $e_\omega = 0.1098\%$. The speed error decreases monotonically over this sequence. The smaller nonmonotonic changes of frequency after $N_e = 8$ remain below 0.11% and reflect the local root correction rather than a change of controlling branch.

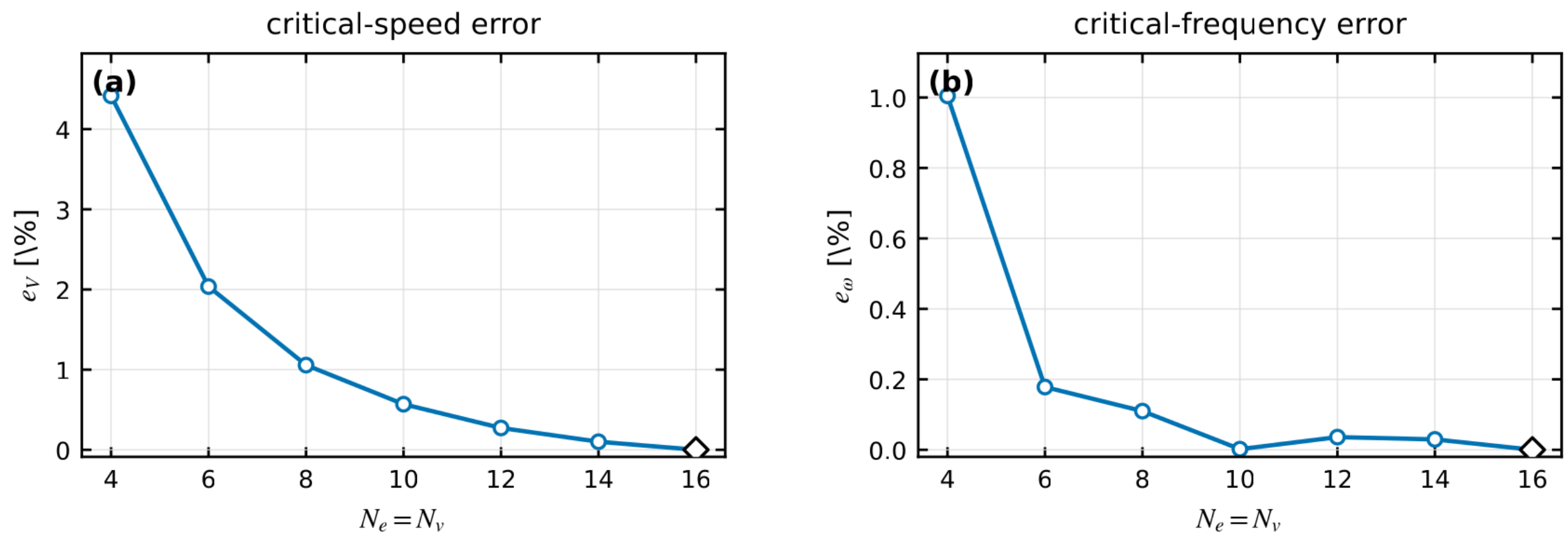

**Fig. 3. Aligned structural-grid refinement relative to the finest common spatial reference.**

For the wake-length study in Fig. 4, the analysis fixes $N_e = N_v = 16$ and $\Delta x_w/c = 0.15$, then varies $L_w/c$ from 7.2 to 9.6 in increments of 0.3. This corresponds to $N_w = 48,50,\ldots,64$. At the baseline wake length, the separate one-dimensional differences are 0.3098% in critical speed and 0.4088% in critical frequency. Both decrease smoothly as the modeled wake length increases.

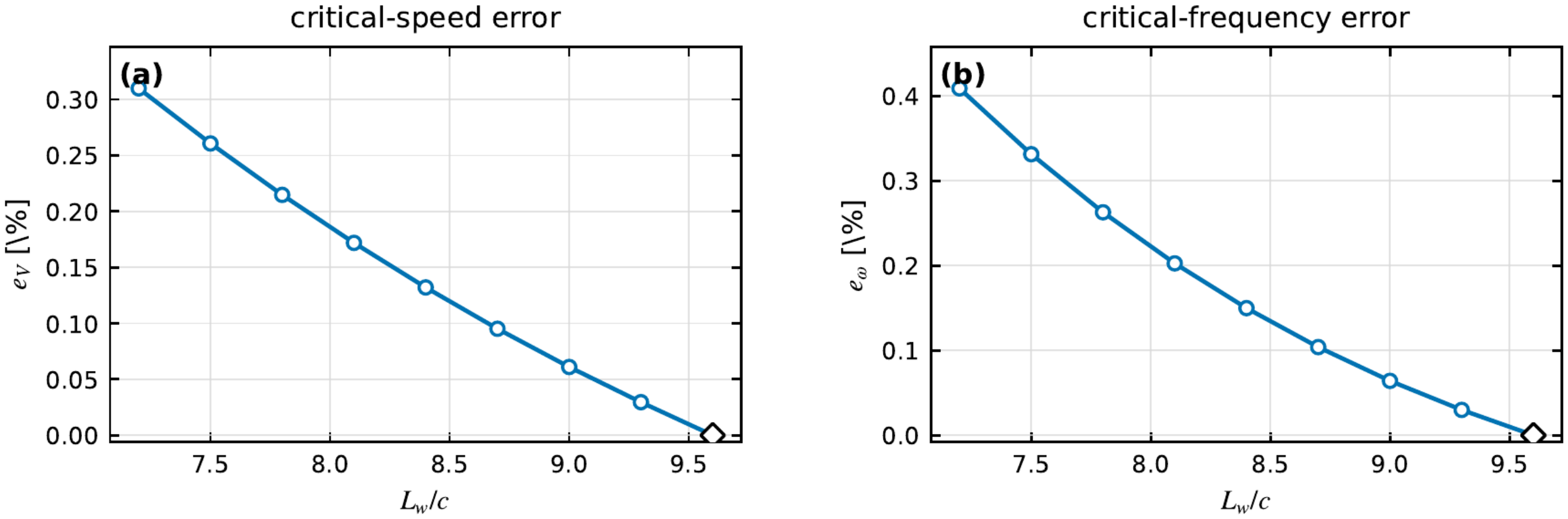

**Fig. 4. Wake-length refinement relative to the longest-wake common spatial reference.**

For the wake-spacing study in Fig. 5, the analysis fixes the finest structural/spanwise grid and the longest wake, then varies the wake spacing through an integer number of rows. The plotted spacings are 0.3000, 0.2667, 0.2400, 0.2087, 0.1811, and 0.1500. The baseline spacing differs from the finest reference by 0.00766% in critical speed and 0.01373% in critical frequency; the intermediate points show a small nonmonotonic variation.

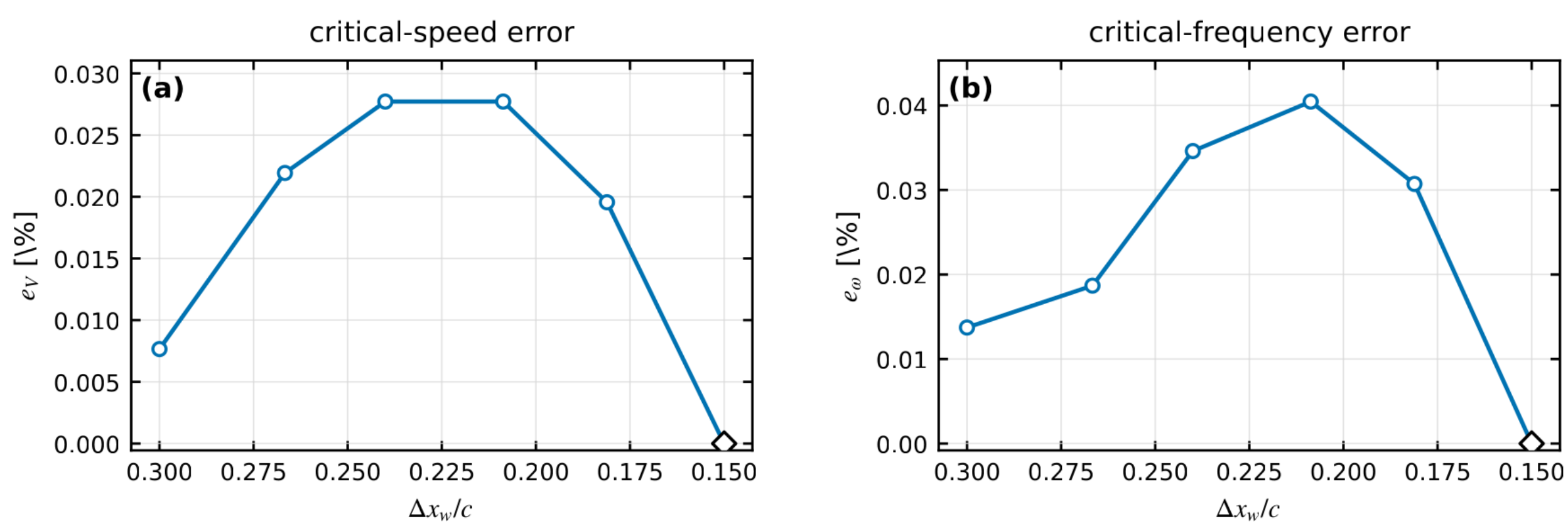

**Fig. 5. Wake-spacing refinement relative to the finest common spatial reference.**

The three curves are one-dimensional refinements that share the same most expensive reference corner; all finite results are shown in Figs. 3–5.

### C. Structural and wake reduction

Let $\Phi_s$ contain retained structural displacement vectors and $\Phi_w$ contain retained transported-wake circulation vectors. The structural displacement and wake circulation are represented by

$$q \simeq \Phi_s a, \qquad \Gamma_w \simeq \Phi_w z_w \tag{26}$$

The ordered baseline vectors and nested retained spaces are

$$(\varphi_1, \dots, \varphi_9) = (\mathrm{IP}_1, \mathrm{B}_1, \mathrm{T}_1, \mathrm{IP}_2, \mathrm{B}_2, \mathrm{T}_2, \mathrm{IP}_3, \mathrm{B}_3, \mathrm{T}_3), \qquad \mathcal{V}_s^{(r_s)} = \mathrm{span}\{\varphi_1, \dots, \varphi_{r_s}\}, \quad \mathcal{V}_s^{(3)} \subset \cdots \subset \mathcal{V}_s^{(9)} \tag{27}$$

Here IP, B, and T denote component-local in-plane bending, out-of-plane bending, and physical low-frequency torsion, respectively; the subscript identifies the component whose local deformation is dominant. The torsional vectors describe elastic section twist and are distinct from the two joint coordinates. For each $r_s = 3,4, \dots, 9$, the first $r_s$ ordered vectors define the retained subspace.

The ROM checks use the fixed baseline analysis grid $N_e = N_v = 8$, $N_u = 6$, $N_w = 24$, $\Delta x_w/c = 0.30$, and $L_w/c = 7.2$, rather than the expensive spatial reference in Eq. (24). Three points sample the parameter range:

$$P1 = (-0.20, 30^\circ), \qquad P2 = (-0.35, 60^\circ), \qquad P3 = (-0.10, 45^\circ) \tag{28}$$

At each point, $(r_s, r_w) = (9,8)$ is computed once and serves as the common ROM reference. The structural sequence varies $r_s = 3{:}\,9$ at $r_w = 8$; the wake sequence varies $r_w = 3{:}\,8$ at $r_s = 9$. These are independent one-parameter checks that share the same reference; they are not a Cartesian product of structural and wake orders.

In addition to Eq. (25), the total aerodynamic transfer error is

$$e_G = 100 \left[ \frac{\sum_k w_k \parallel G_r(i\Omega_k) - G_{ref}(i\Omega_k) \parallel_F^2}{\sum_k w_k \parallel G_{ref}(i\Omega_k) \parallel_F^2} \right]^{1/2} \% \tag{29}$$

where the same frequency grid, port weights, direct term, and pressure-rate term are used at every order. Figures 6 and 7 report relative critical-speed error, relative critical-frequency error, and the total FRF error so that stability and input–output effects are visible on their own scales.

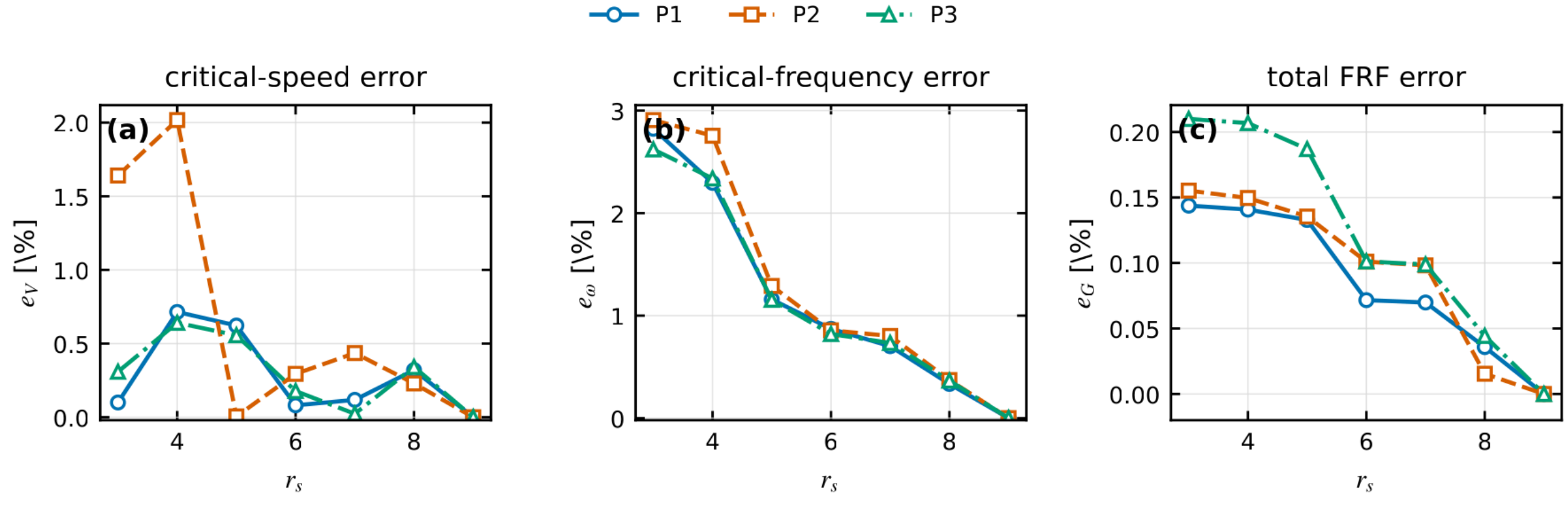

**Fig. 6. Structural reduced-order convergence at three parameter points on the baseline analysis grid.**

The structural sequence is not strictly monotone in critical-speed error because a truncated basis perturbs both candidate roots before the lower envelope is selected. Across all displayed orders and points, the largest critical-speed error is 2.02%, the largest critical-frequency error is 2.91%, and the largest total FRF error is 0.210%. The baseline structural basis uses $r_s = 9$.

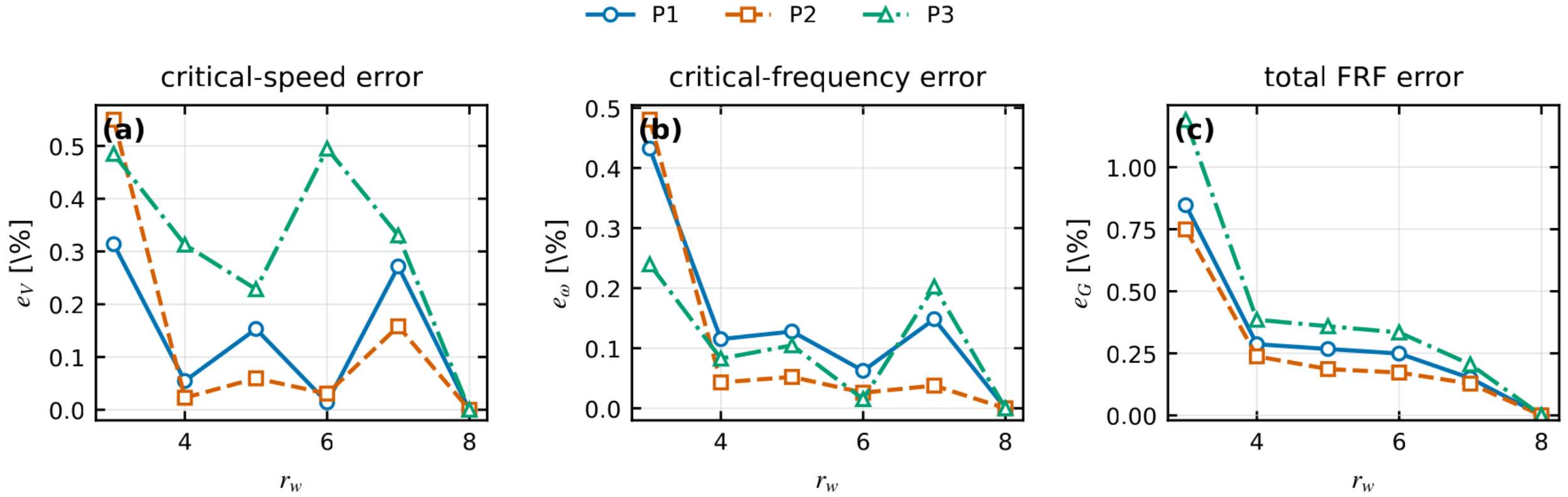

**Fig. 7. Wake reduced-order convergence at three parameter points on the baseline analysis grid.**

At the baseline wake order $r_w = 4$, the P1–P3 critical-speed errors range from 0.0228% to 0.3129%, the critical-frequency errors from 0.0431% to 0.1152%, and the total FRF errors from 0.2377% to 0.3851%. Increasing wake order reduces the total FRF error overall, while the stability errors remain small and mildly nonmonotone. This distinction is expected because the flutter boundary depends on a few coupled roots whereas Eq. (29) samples the complete retained transfer over frequency.

**Table 3. Representative refinement and baseline-order differences.**

| Comparison | Critical-speed error [%] | Critical-frequency error [%] | Total FRF error [%] |
|---|---|---|---|
| Structural grid, $N_e = N_v = 8$ vs. common reference | 1.0566 | 0.1098 | – |
| Wake length, $L_w/c = 7.2$ vs. common reference | 0.3098 | 0.4088 | – |
| Wake spacing, $\Delta x_w/c = 0.30$ vs. common reference | 0.00766 | 0.01373 | – |
| Structural ROM, all $r_s = 3{:}\,9$, P1–P3 | 0 to 2.017 | 0 to 2.907 | 0 to 0.210 |
| Baseline wake ROM, $r_w = 4$, P1–P3 | 0.0228 to 0.3129 | 0.0431 to 0.1152 | 0.2377 to 0.3851 |

The baseline analysis grid and reduced orders defined above are used in the remaining calculations. The finest spatial corner and highest ROM orders serve as the corresponding convergence references.

## IV. Linear Flutter and Modal Competition

### A. Flutter boundary and the critical-frequency valley

At each scheduled geometry and speed, the generalized eigenproblem is

$$L(\mu)v_j = \lambda_j E(\mu)v_j, \qquad \mu = (\epsilon_A, \phi, V^*), \qquad \lambda_j = \alpha_j + i\omega_j \tag{30}$$

The L and H labels denote the two oscillatory pairs continued through the parameter region and distinguished by their low and high frequencies; they are not reassigned by instantaneous frequency sorting. A branch-specific neutral speed $V_{j,0}^*$ satisfies $\alpha_j\left(V_{j,0}^*\right) = 0$. The *controlling branch* is the one that attains the minimum neutral speed at a stated configuration, and the flutter envelope is

$$V_f^*(\epsilon_A, \phi) = \min_j V_{j,0}^*(\epsilon_A, \phi), \qquad f_f^* = \frac{|\omega_f|}{\omega_0} \tag{31}$$

For a simple neutral branch, let $g_j = \mathrm{Re}\lambda_j$ and assume $\partial_{V^*} g_j \neq 0$. The branch tangent and the associated frequency action follow from

$$\frac{\partial V_{j,0}^*}{\partial \mu_a} = -\frac{\partial_{\mu_a} g_j}{\partial_{V^*} g_j}, \qquad \frac{d\omega_{j,0}}{d\mu_a} = \partial_{\mu_a} \mathrm{Im}\lambda_j + \partial_{V^*} \mathrm{Im}\lambda_j \frac{\partial V_{j,0}^*}{\partial \mu_a}, \quad \mu_a \in \{\epsilon_A, \phi\}$$

This first-order information distinguishes a branch action from a sampled change of the scalar envelope and can support local stability-margin and configuration-design reasoning. Section V constructs the required descriptor derivatives; no optimization calculation is used in the present results.

Figure 8 shows four representative fold-angle slices from the full two-parameter flutter analysis. All 78 neutral solutions are finite. Critical speed changes smoothly on the displayed angular grid, whereas the critical normalized frequency at the more negative offsets follows the H branch near 0.29, the L branch near 0.15, and then H again. Figures G1 and G2 contain the complete six-offset set.

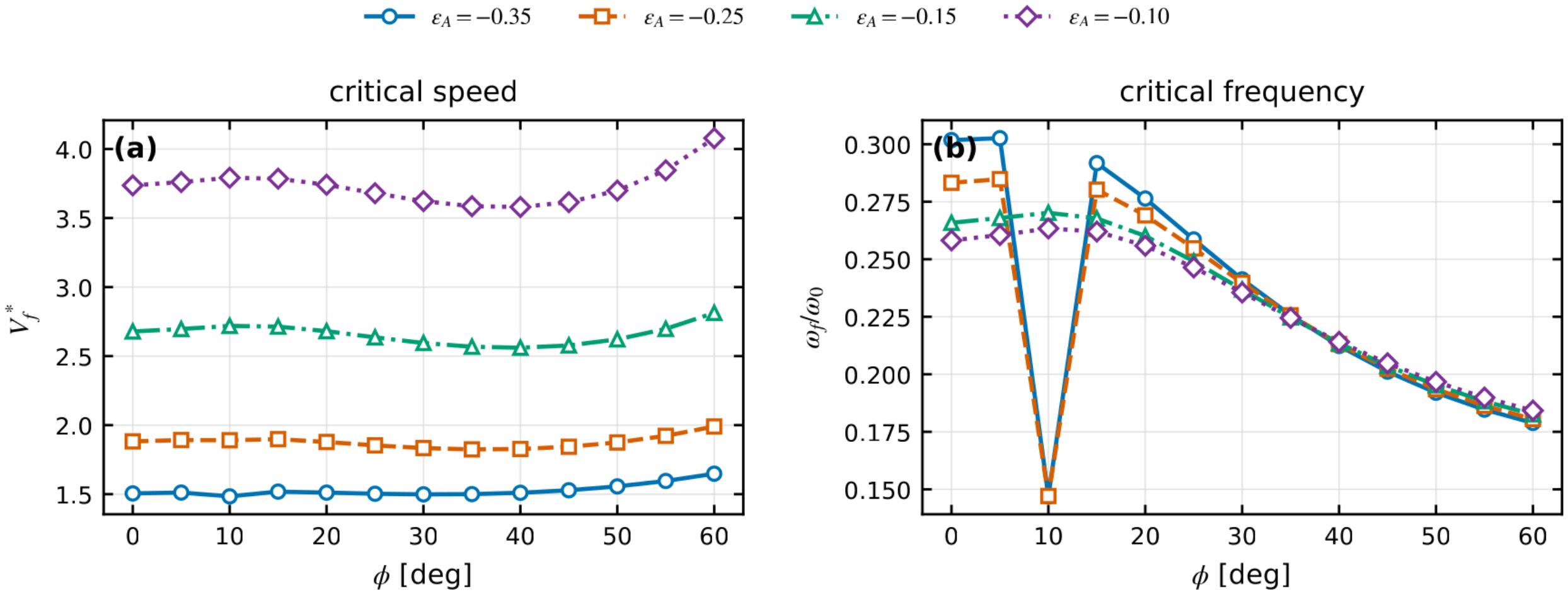

**Fig. 8. Flutter speed and critical-frequency slices for four structural-axis offsets.**

Continuous branch tracking shows that the frequency valley occurs when the L and H neutral branches alternately attain the minimum in Eq. (31). A scalar flutter map records only the controlling root; the following root-locus analysis retains both tracked pairs.

**B. Representative H–L–H crossing sequence**

The -0.25 structural-axis-offset slice contains two changes of the controlling branch within a narrow fold-angle interval. Figure 9 displays the full positive-imaginary root histories at three fold angles, with local insets centered on the first neutral crossing. The controlling branch is H at 6.25° (normalized critical frequency 0.295277), L at 12.5° (0.147218), and H again at 13.125° (0.293424). The global trajectories show each tracked pair within the surrounding spectrum, while the insets resolve the crossing order.

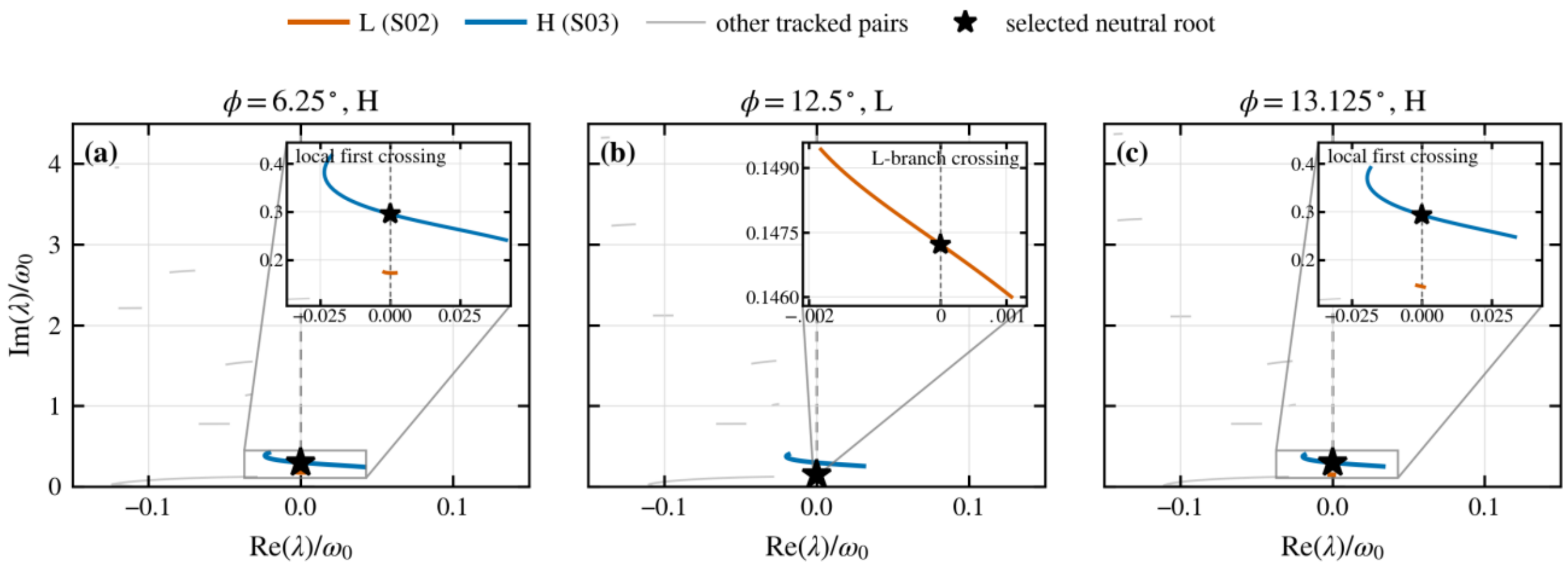

**Fig. 9. Tracked root loci and first-crossing insets at structural-axis offset -0.25; stars mark the controlling neutral roots.**

For comparison with the customary flutter presentation, define

$$g_j = \frac{Re\lambda_j}{|Im\lambda_j|}, \qquad f_j^* = \frac{|Im\lambda_j|}{\omega_0} \tag{32}$$

With this convention, negative, zero, and positive normalized damping denote stable, neutral, and unstable oscillatory roots, respectively. Figure 10 applies Eq. (32) to the same three configurations. The damping panels resolve the neutral neighborhoods, and the associated frequency panels show that the L and H branches remain distinct families throughout the speed sweep.

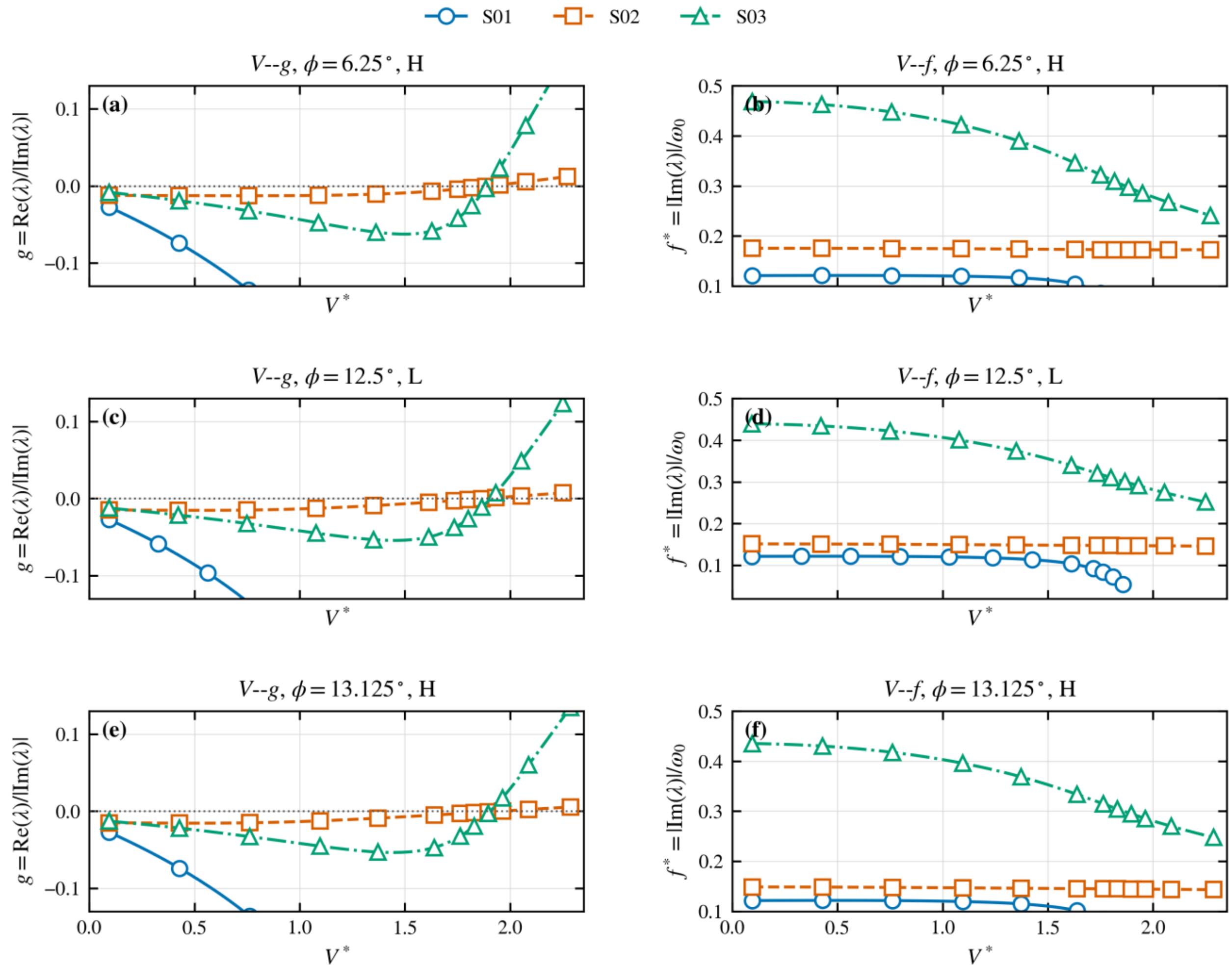

**Fig. 10. V-g and V-f diagrams for the high-low-high sequence at fold angles 6.25, 12.5, and 13.125 deg.; at 12.5 deg., S01 is shown through the first flutter speed.**

Figure 9 establishes the crossing order in complex-eigenvalue space, and Fig. 10 follows the same L and H families through normalized damping and frequency. Together they show that alternating control of the lower neutral-speed envelope produces the frequency valley.

### C. Component participation, structural energy, and ancestor shapes

Because the coupled descriptor pencil is non-self-adjoint, a right-eigenvector norm alone does not measure how a state block participates in an eigenvalue. Let $u_j = E^H w_j$, so that $u_j^H v_j = w_j^H E v_j = 1$, and partition the 26 states into

$$\mathcal{G} = \{B_1, B_2, B_3, J_{12}, J_{23}, \text{wake}\}.$$

The block contraction and normalized left–right participation are [36]

$$\chi_{j,g} = u_{j,g}^H v_{j,g}, \qquad p_{j,g} = \frac{|\chi_{j,g}|}{\sum_{h\in\mathcal{G}} |\chi_{j,h}|} \tag{33}$$

An independent mechanical measure is obtained from the displacement part of the mode. With $\mathcal{U}_{j,g}$ denoting the component beam elastic strain energy or the corresponding hinge potential energy,

$$e_{j,g} = \frac{\mathcal{U}_{j,g}}{\sum_{h\in\mathcal{G}_s} \mathcal{U}_{j,h}}, \qquad \mathcal{G}_s = \{B_1, B_2, B_3, J_{12}, J_{23}\} \tag{34}$$

Figure 11 reports both definitions for the controlling root at the three crossings. The L-controlled case concentrates participation and elastic energy in $B_1$, whereas the two H-controlled cases distribute more of both measures through $B_2$. The individual hinge-energy shares remain at or below 1.3%. The left–right and energy measures therefore agree on the principal $B_1$–$B_2$ redistribution while measuring different properties of the mode.

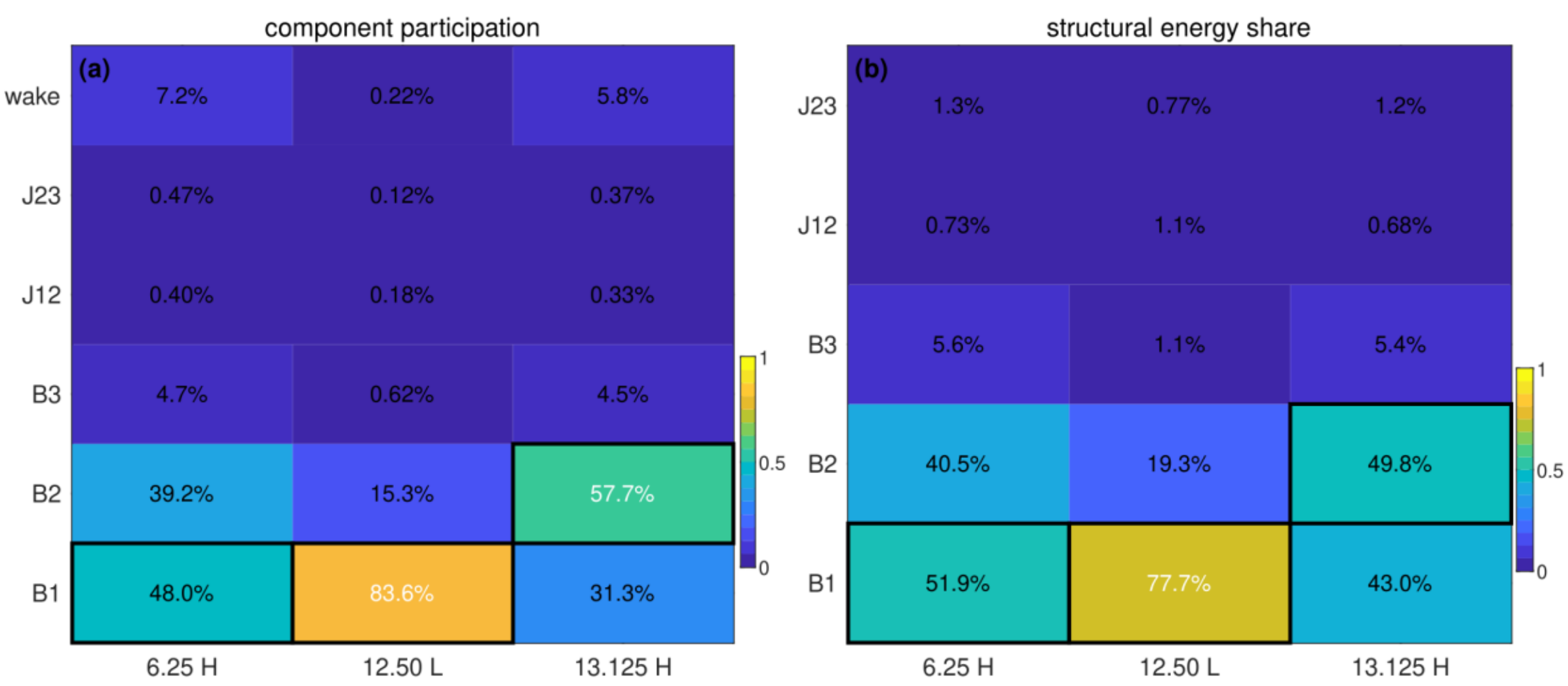

**Fig. 11. Left-right participation and structural-energy shares of the controlling neutral roots across the high-low-high sequence.**

The geometric comparison uses the low-aerodynamic-loading ancestors at the valley configuration, where the coupled shapes remain close to their structural progenitors. A single phase factor is fixed for each L/H ancestor before the state is returned to component-root coordinates. For Fig. 12, each component is rendered as a continuous rectangular beam; element color gives elastic strain-energy density on a scale shared across each row, and a light wireframe shows the attached aerodynamic surface. The displayed geometry is independently amplified within each panel so that bending and twist remain visually distinguishable; Fig. 11 provides the quantitative comparison of component contribution.

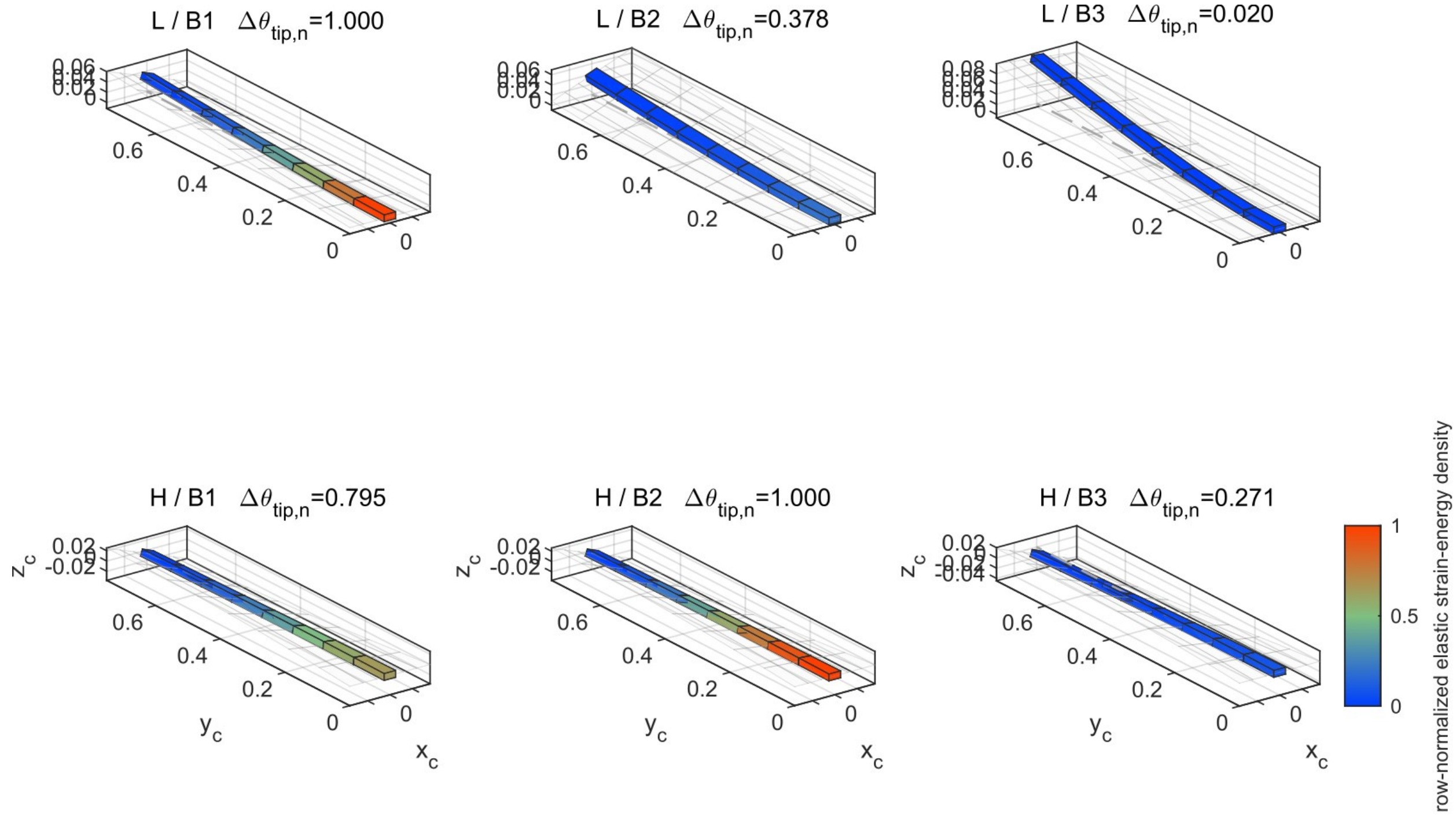

**Fig. 12. Component-local low- and high-frequency ancestor shapes at the valley configuration; color denotes normalized elastic strain-energy density.**

The L ancestor concentrates its tip-rotation increment and strain energy in B1, with rapid attenuation through B2 and B3. The H ancestor distributes the response more evenly between B1 and B2 and retains a clearer B3 deformation. Together, Figs. 11 and 12 link the change of controlling branch to a component-level redistribution of structural deformation and coupled structural–wake participation.

Fold angle changes the relative surface–wake geometry and the projection of each component velocity onto the local aerodynamic normal, whereas $\epsilon_A$ changes attachment-velocity and pressure-load lever arms. These distinct parameter paths alter two continuous structural–wake branches, which alternately form the lower envelope twice on the representative slice and produce the observed critical-frequency valley.

## V. Double-Hopf Localization and 1:2 Normal Form

### A. Differentiable local descriptor pencil

The local continuation model uses one fixed Petrov–Galerkin wake coordinate over each parameter band. If $V_w, W_w$ contain the common right and left wake directions, then

$$W_w^H V_w = I_4, \qquad A_{1r} = W_w^H A_1 V_w, \qquad B_{1r} = W_w^H B_1, \qquad C_{1r} = C_1 V_w \tag{35}$$

The common basis fixes the wake-state gauge; it is not itself fitted. Only $A_1, B_1, C_1$ are transformed by Eq. (35). The direct, pressure-rate, normal-flow, and structural blocks remain in their natural structural-input/aerodynamic-output coordinates [19, 21]. The configuration-dependent primitive set is

$$\mathcal{P} = \{A_{1r}, B_{1r}, C_{1r}, D_1, \dot{D}_1, N_{q1}, N_v, M_s, C_s, K_s\}$$

Here $\dot{D}_1$ denotes the pressure-rate block, not the time derivative of $D_1$. For normalized local coordinates $\hat{\epsilon} = (\epsilon_A - \epsilon_0)/0.05$ and $\hat{\phi} = (\phi - \phi_0)/(2.5°)$, each $P \in \mathcal{P}$ is represented in the common total-degree-two chart

$$\hat{P}(\hat{\epsilon}, \hat{\phi}) = P_{00} + P_{10}\hat{\epsilon} + P_{01}\hat{\phi} + P_{20}\hat{\epsilon}^2 + P_{11}\hat{\epsilon}\hat{\phi} + P_{02}\hat{\phi}^2, \qquad P \in \mathcal{P} \tag{36}$$

Constant and affine dependences are lower-degree members of the same chart. The speed factors in Eq. (19) and the algebraic assembly in Eqs. (20)–(21) remain exact, defining the structured map

$$(\mathcal{P}, U) \mapsto (E, L)$$

Using $M_e, C_e, K_e$ from Eq. (20), the local common-coordinate pencil is assembled as

$$E = \mathrm{blkdiag}(I, M_e, I_4), \qquad L = \begin{bmatrix} 0 & I & 0 \\ -K_e & -C_e & UC_{1r} \\ U^2 B_{1r} N_{q1} & U B_{1r} N_v & U A_{1r} \end{bmatrix}$$

This block separation preserves the origin of each parameter action. Fold angle changes relative surface–wake geometry, component orientation, and normal-flow projection. In the present chord-axis-fold geometry, a common chordwise translation associated with $\epsilon_A$ does not change the relative vortex–control-point influence geometry, but the offset does change attachment-velocity and pressure-load moment arms. The two configuration parameters therefore act through distinct primitive pathways. The uniform local chart is applied to all members of $\mathcal{P}$, with invariant and affine actions represented as lower-degree cases.

The resulting derivative construction is blockwise: $\partial_{\mu_a}\mathcal{P}$ generates $(\partial_{\mu_a}E, \partial_{\mu_a}L)$, which then gives the eigenvalue action. For a simple generalized eigenvalue normalized by Eq. (22) [22],

$$\frac{\partial \lambda_j}{\partial \mu_a} = w_j^H \left( \frac{\partial L}{\partial \mu_a} - \lambda_j \frac{\partial E}{\partial \mu_a} \right) v_j \tag{37}$$

This organization provides a systematic route to residual, descriptor-matrix, and eigenvalue actions. Exact speed actions are retained analytically; the configuration actions used here are obtained by differentiating the local primitive chart in fixed coordinates.

Let $\lambda_L = \beta_L + i\omega_L$ and $\lambda_H = \beta_H + i\omega_H$ denote the two branches identified in Sec. IV. Simultaneous neutrality and 1:2 resonance require

$$\beta_L = 0, \qquad \beta_H = 0, \qquad \Delta_{12} = \frac{\omega_H - 2\omega_L}{\omega_0} = 0 \tag{38}$$

Continuation of the first two equations produces the low- and high-angle double-Hopf curves in Fig. 13(a). Along the high-angle curve, the two normalized frequency branches approach a 1:2 ratio and the detuning crosses zero, Figs. 13(b) and 13(c). Solving all three conditions gives, to a precision commensurate with the spatial, reduced-order, and local descriptor models,

$$(\epsilon_A^*, \phi^*, V^*) = (-0.2430,\ 12.65°,\ 1.943) \tag{39}$$

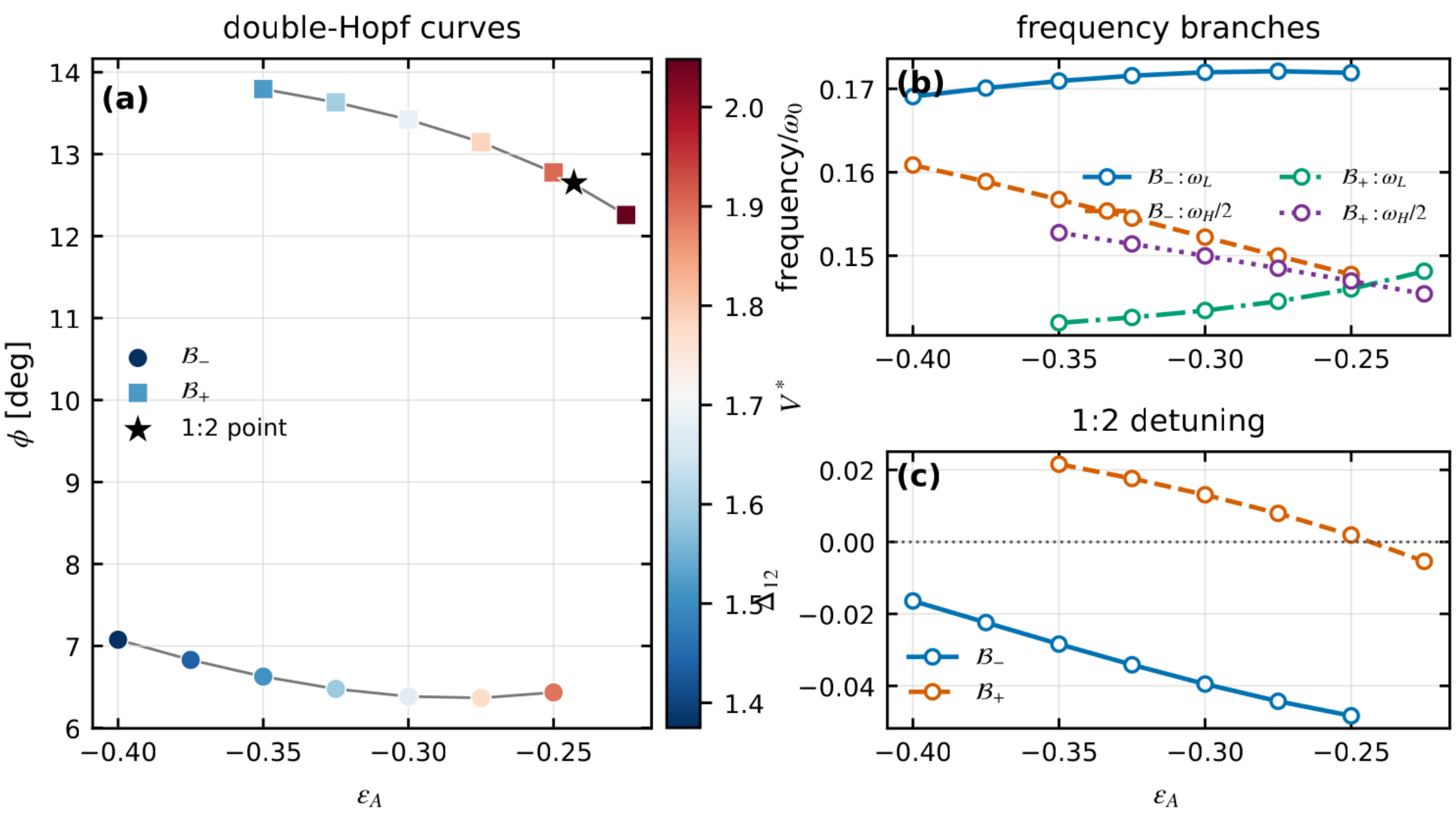

**Fig. 13. Localized resonant double-Hopf point: a) simultaneous-neutrality curves; b) frequency branches; c) 1:2 detuning.**

**Table 4. Linear quantities from the local descriptor at the resonant double-Hopf point.**

| Quantity | Value |
|---|---|
| Dimensional speed $U^*$ | 4.945 m s$^{-1}$ |

| Quantity | Value |
|---|---|
| Growth rates $(\beta_L, \beta_H)$ | $(3.13 \times 10^{-7}, -1.63 \times 10^{-6})$ s$^{-1}$ |
| Frequencies $(\omega_L, \omega_H)$ | (4.143, 8.287) rad s$^{-1}$ |
| Normalized frequencies | (0.1465, 0.2931) |
| Normalized detuning $\Delta_{12}$ | $1.47 \times 10^{-7}$ |

The small growth-rate and detuning values in Table 4 are the residuals obtained when the three-condition solution is re-evaluated in the coupled descriptor model. The two critical pairs are simple, the remaining spectrum is separated from the imaginary axis in the local calculation, and the two quadratic resonant coefficients reported below are nonzero. Together with the unfolding rank discussed after Eq. (45), these are the nondegeneracy conditions used to interpret the local double-Hopf reduction.

### B. Center-manifold construction and coordinate scaling

At the resonant scheduling point, the local 26-state reduced model retains the quadratic and cubic structural restoring-force terms:

$$E\dot{x} = Lx + F_2(x,x) + F_3(x,x,x) + \mathcal{O}(\| x \|^4), \qquad x \in \mathbb{R}^{26} \tag{40}$$

Here $F_2$ and $F_3$ are obtained from component-wise static condensation of the structural restoring force. The aerodynamic and inertial operators are evaluated and held fixed at the scheduling point. Accordingly, Eq. (40) is the nonlinear model adopted for the local study, not a complete nonlinear-aerodynamic or configuration-dependent-inertia system. Let the critical right and adjoint-left eigenspace bases be

$$V_c = [v_L, \bar{v}_L, v_H, \bar{v}_H], \qquad W_c = [w_L, \bar{w}_L, w_H, \bar{w}_H], \qquad W_c^H E V_c = I_4$$

and let

$$\zeta = (\zeta_L, \bar{\zeta}_L, \zeta_H, \bar{\zeta}_H)^T \in \mathbb{C}^4$$

The third-order parameterization and reduced vector field are constructed together within the standard nonlinear-oscillation, bifurcation, and invariant-manifold framework [37–42],

$$x = \mathcal{K}^{[3]}(\zeta) = V_c\zeta + K_2(\zeta) + K_3(\zeta), \qquad \dot{\zeta} = N^{[3]}(\zeta) \tag{41a}$$

$$E\, D\mathcal{K}^{[3]}(\zeta) N^{[3]}(\zeta) = L\mathcal{K}^{[3]}(\zeta) + F_2(\mathcal{K},\mathcal{K}) + F_3(\mathcal{K},\mathcal{K},\mathcal{K}) + \mathcal{O}(\| \zeta \|^4) \tag{41b}$$

The stored center eigenvectors define the phase origin at the resonant point. Neighboring eigenvectors are aligned by requiring their biorthogonal overlap with the preceding vector to be positive real, while maintaining $W^H EV = I$. All complex coefficients and phase plots below use this gauge. This declaration is essential because independent rephasing of $v_L$ and $v_H$ changes the phases of the resonant coefficients without changing the physical dynamics.

The displayed normal form uses the amplitude and time transformations

$$z_j = \frac{\zeta_j}{\rho_{work}}, \qquad \tau = \omega_0 t, \qquad \rho_{work} = 0.0139247665, \qquad \omega_0 = 28.2743339 \, \mathrm{rad\, s^{-1}} \tag{42}$$

If a dimensional monomial coefficient $c_\alpha$ has total degree $p$, then

$$\hat{c}_\alpha = c_\alpha \frac{\rho_{work}^{p-1}}{\omega_0} \tag{43}$$

Equation (43) gives the explicit relation between dimensional center coordinates and the dimensionless coefficients reported here. The reported coefficient magnitudes are consequently much smaller than coefficients written in unscaled amplitude and dimensional time; this difference is a coordinate effect, not a change in the underlying vector field.

Retaining the two 1:2 resonant quadratic monomials and the four cubic saturation monomials [26] gives

$$z_L' = \left(\hat{\beta}_L + i\hat{\omega}_L\right) z_L + \hat{c}\, \bar{z}_L z_H + \hat{a}_{LL} z_L |z_L|^2 + \hat{a}_{LH} z_L |z_H|^2 \tag{44a}$$

$$z_H' = \left(\hat{\beta}_H + i\hat{\omega}_H\right) z_H + \hat{d}\, z_L^2 + \hat{a}_{HL} z_H |z_L|^2 + \hat{a}_{HH} z_H |z_H|^2, \qquad ' = d/d\tau \tag{44b}$$

The terms $\hat{c}\bar{z}_L z_H$ and $\hat{d} z_L^2$ are the reciprocal 1:2 transfer channels. The two coefficients depend on both the structural nonlinear tensors and their projection through the configuration-dependent critical right and adjoint left eigenvectors. Unless stated otherwise, the coefficients and integrations below use $\beta = 0.25$, where $\beta$ is the dimensionless multiplier of the reference cubic hinge coefficient $\bar{k}_{\theta_j}^{(3)}$ with units $\mathrm{N\, m\, rad^{-3}}$; Sec. VII defines the hinge law. Figure 14 separates magnitude and phase so that coefficients spanning more than two orders of magnitude remain readable. Table 5 gives the numerical values used in all subsequent integrations.

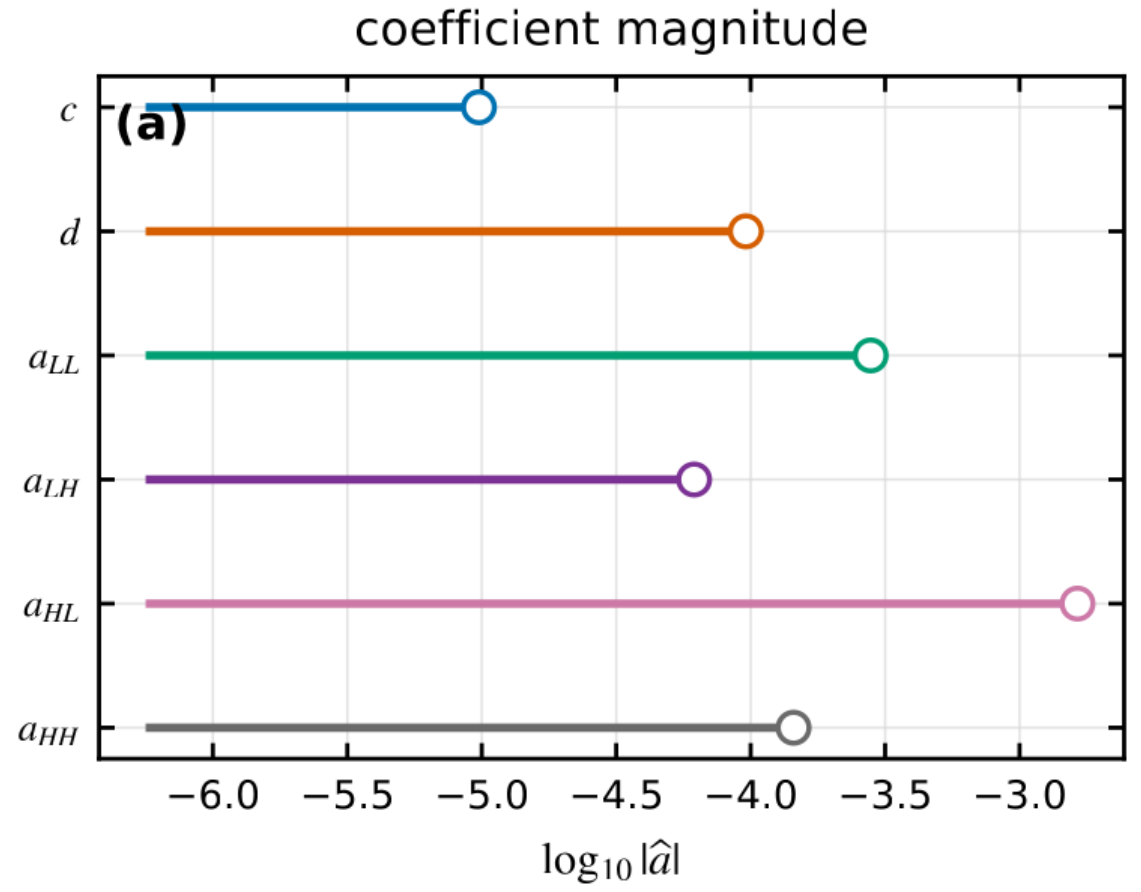

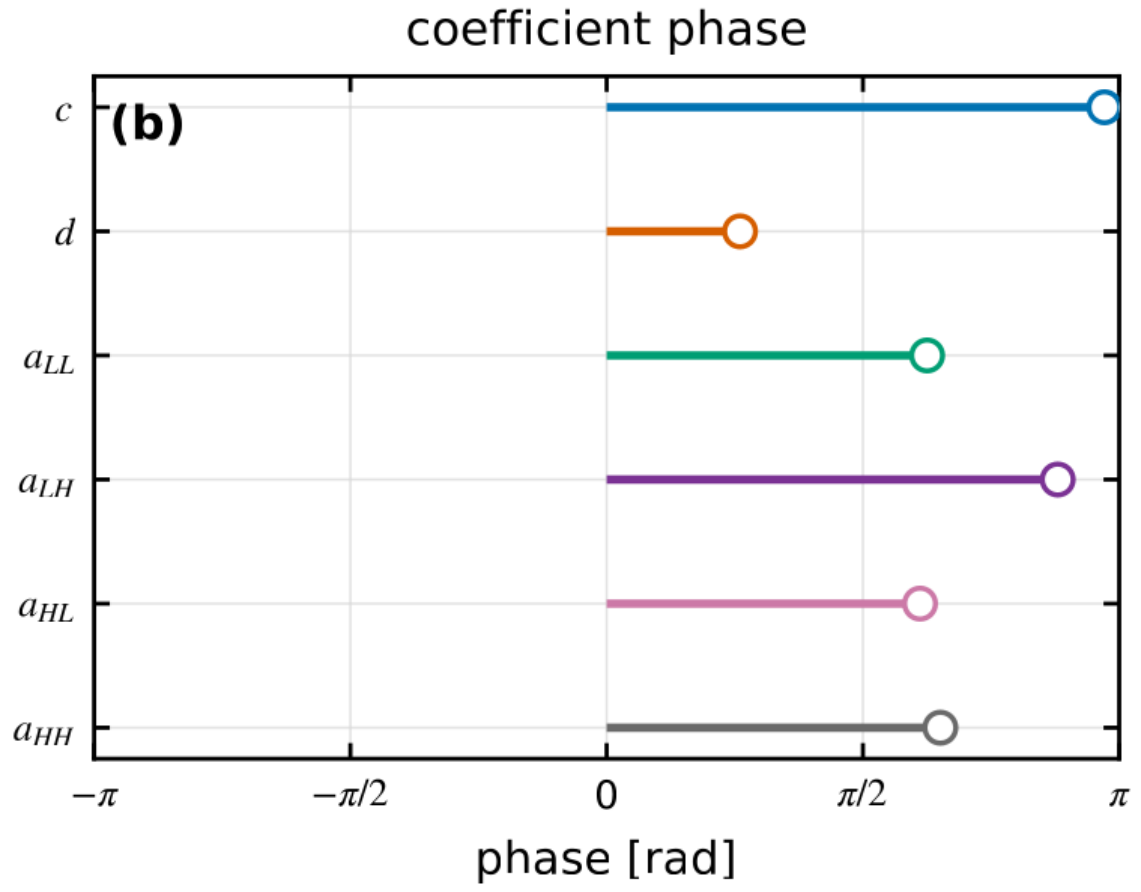

**Fig. 14. Dimensionless cubic normal-form coefficients: a) logarithmic magnitudes; b) phases.**

**Table 5. Complex dimensionless coefficients in Eqs. (44a)–(44b) at β = 0.25.**

| Coefficient | Real part | Imaginary part |
|---|---|---|
| $\hat{c}$ | $-9.72256 \times 10^{-6}$ | $8.82713 \times 10^{-7}$ |
| $\hat{d}$ | $6.56222 \times 10^{-5}$ | $7.00013 \times 10^{-5}$ |
| $\hat{a}_{LL}$ | $-1.07156 \times 10^{-4}$ | $2.58105 \times 10^{-4}$ |
| $\hat{a}_{LH}$ | $-5.73001 \times 10^{-5}$ | $2.26682 \times 10^{-5}$ |
| $\hat{a}_{HL}$ | $-5.63422 \times 10^{-4}$ | $1.54298 \times 10^{-3}$ |
| $\hat{a}_{HH}$ | $-6.60319 \times 10^{-5}$ | $1.28463 \times 10^{-4}$ |

### C. Reachable unfolding and representative dynamics

The physical parameters unfold the two growth rates and detuning through

$$\mathcal{U}(\mu) = [\hat{\beta}_L \quad \hat{\beta}_H \quad \hat{\delta}]^T, \qquad \hat{\delta} = \frac{\omega_H - 2\omega_L}{\omega_0}, \qquad J_{\mathcal{U}} = \frac{\partial \mathcal{U}}{\partial(\epsilon_A, \phi, V^*)} \tag{45}$$

The singular values of the raw Jacobian are 22.7600, 6.30060, and 0.025411; after scaling its columns by the local parameter-box widths, they are 0.371058, 0.100190, and 0.002538. The scaled map is locally full rank but strongly anisotropic, with condition number approximately 146; the weakest input combination has much less authority than the two dominant growth-rate combinations. Figure 15 shows the local parameter samples and the resonant point.

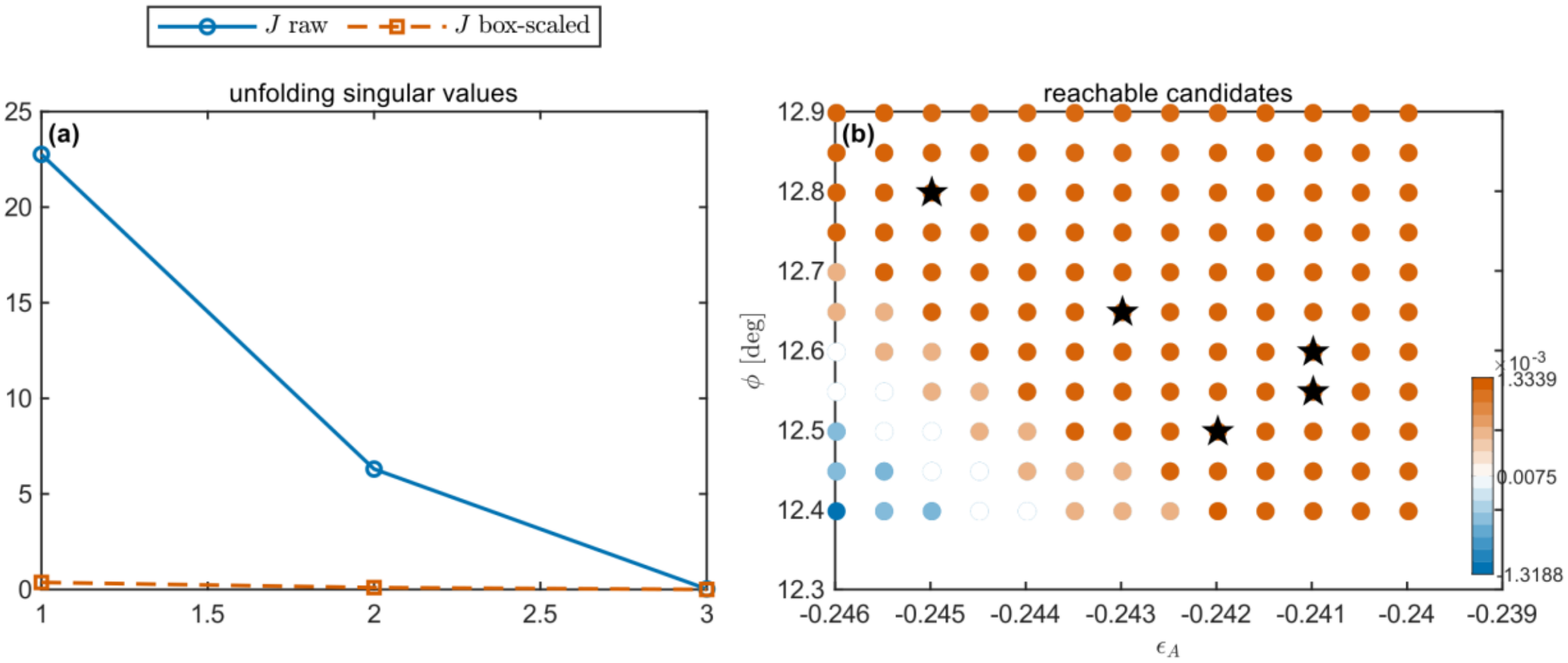

**Fig. 15. Local resonant unfolding: a) scaled singular values; b) parameter samples.**

Writing $z_L = r_L e^{i\theta_L}$, $z_H = r_H e^{i\theta_H}$, and $\psi = \theta_H - 2\theta_L$ gives the quotient dynamics

$$r_L{}' = r_L\left[\hat{\beta}_L + \mathrm{Re}(\hat{a}_{LL})r_L^2 + \mathrm{Re}(\hat{a}_{LH})r_H^2 + r_H\mathrm{Re}\left(\hat{c}e^{i\psi}\right)\right] \tag{46a}$$

$$r_H{}' = r_H\left[\hat{\beta}_H + \mathrm{Re}(\hat{a}_{HL})r_L^2 + \mathrm{Re}(\hat{a}_{HH})r_H^2\right] + r_L^2\mathrm{Re}\left(\hat{d}e^{-i\psi}\right) \tag{46b}$$

$$\begin{aligned} \psi' &= \hat{\delta} + \nu_L r_L^2 + \nu_H r_H^2 + \frac{r_L^2}{r_H}\mathrm{Im}\left(\hat{d}e^{-i\psi}\right) - 2r_H\mathrm{Im}\left(\hat{c}e^{i\psi}\right), \\ \nu_L &= \mathrm{Im}(\hat{a}_{HL}) - 2\mathrm{Im}(\hat{a}_{LL}), \\ \nu_H &= \mathrm{Im}(\hat{a}_{HH}) - 2\mathrm{Im}(\hat{a}_{LH}) \end{aligned} \tag{46c}$$

The complex Cartesian equations are used when $r_H = 0$. At the representative offset point, the trajectory in Fig. 16 evolves through a narrow amplitude region and approaches a phase-locked state. The phase is displayed as $(\cos\psi, \sin\psi)$ on the unit circle; this avoids a branch-cut line that would appear in an unwrapped angle plot.

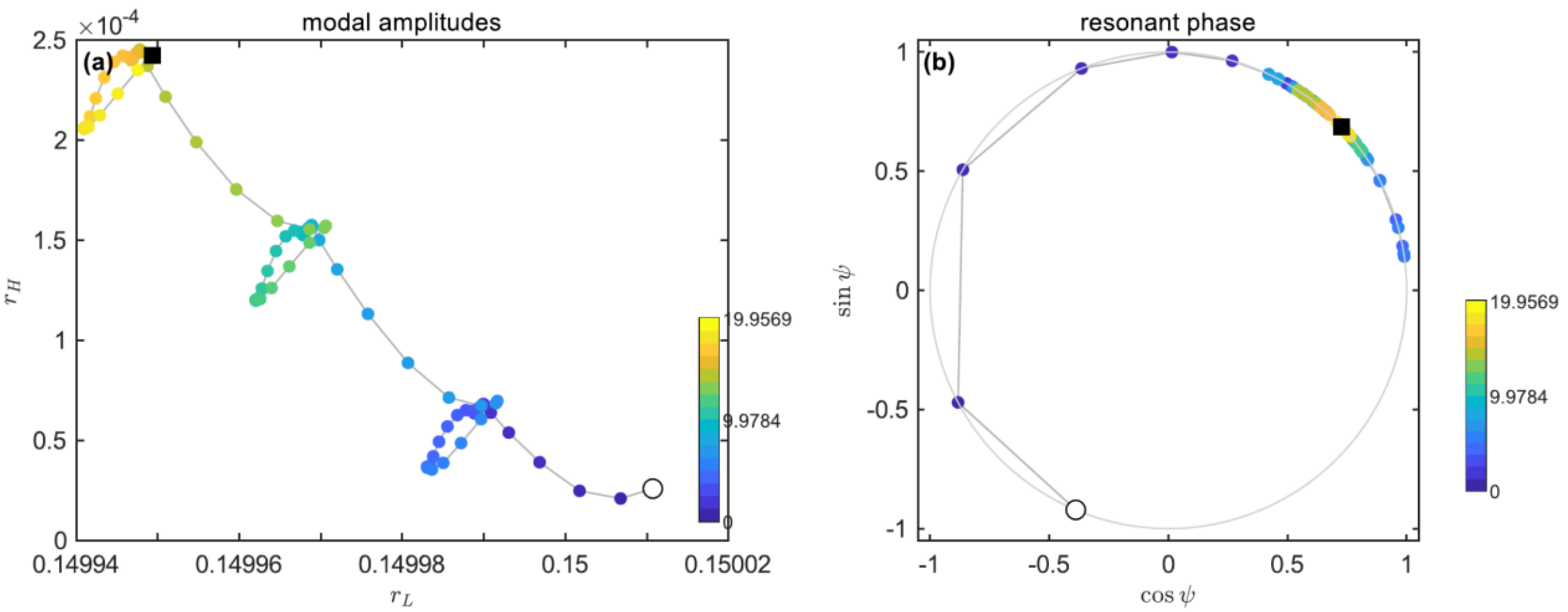

**Fig. 16. Representative normal-form trajectory: a) modal amplitudes; b) resonant phase on the unit circle.**

Figure 17 resolves the same evolution in time. Because the amplitude changes occur on different scales, the low-mode departure from 0.15 and the high-mode amplitude use separate scientific multipliers. The circular coordinates show the slow settling of the resonant phase.

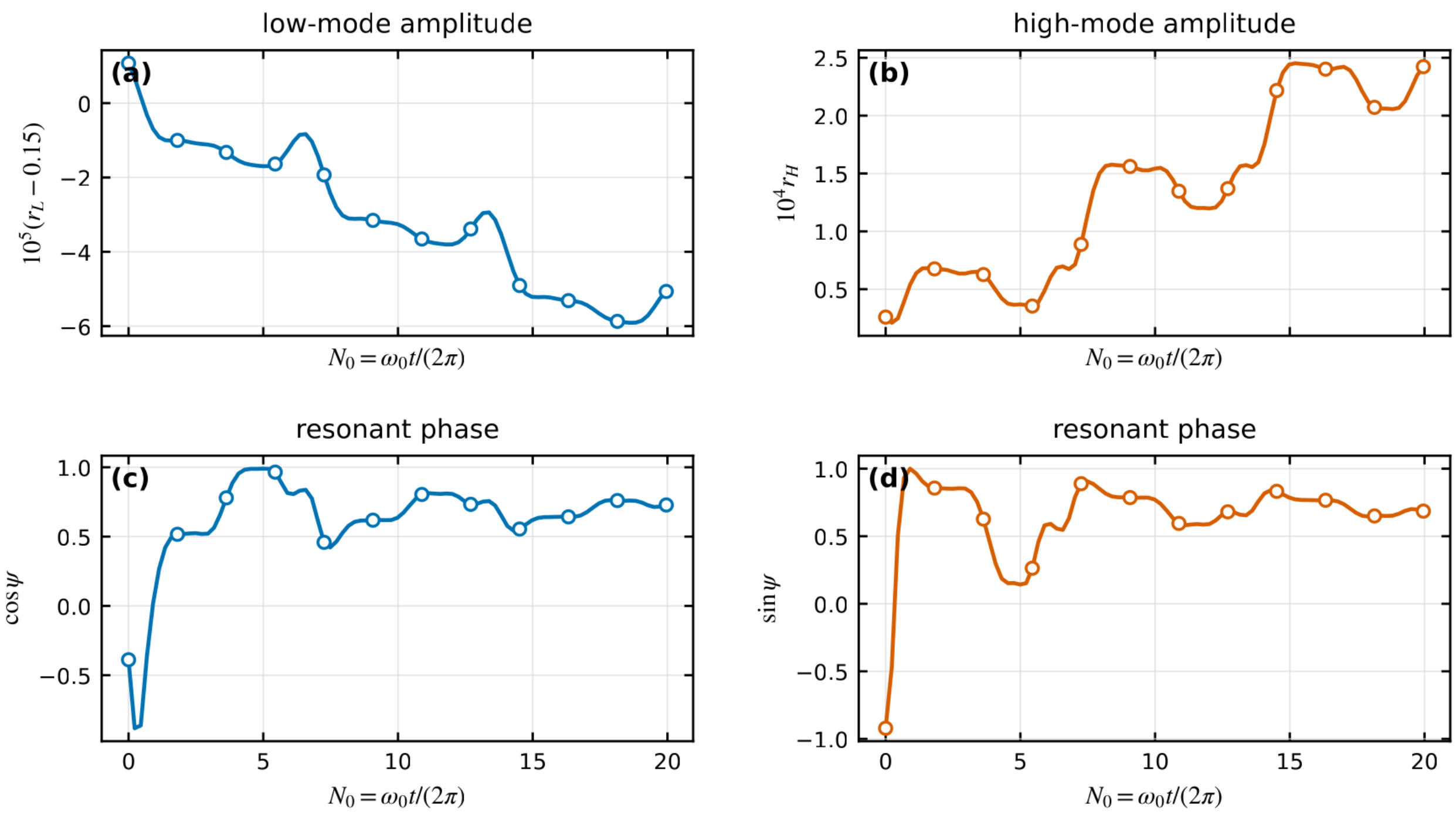

**Fig. 17. Modal amplitudes and resonant phase along the trajectory in Fig. 16.**

## VI. Time-Domain Reconstruction and Periodic Responses

### A. Reconstruction of physical observations

The four-dimensional cubic normal-form solution is returned to the 26-state physical coordinate through the same parameterization used in the invariance equation,

$$z' = \widehat{N}^{[3]}(z), \qquad x_{NF}(t) = \mathcal{K}^{[3]}\big(\rho_{work} z(\omega_0 t)\big), \qquad y_{NF}(t) = C x_{NF}(t) \tag{47}$$

The initial state of the 26-state reduced nonlinear aeroelastic model is set by the identical map,

$$x_{26}(0) = \mathcal{K}^{[3]}\big(\rho_{work} z(0)\big) \tag{48}$$

The observation operator $C$ extracts B3 tip twist and the two physical fold coordinates $\theta_{12}$ and $\theta_{23}$. Figure 18 compares the reconstructed normal-form observations with the 26-state model at the resonant reference point $P_0$ of Eq. (39), and Fig. 19 repeats the consistency check at the representative offset point $P_s$ marked in Fig. 15. Both computations use $\beta = 0.25$ in the hinge law introduced in Sec. VII, the same mapped initial state, and exactly the same 1601 sample times. The left column retains each complete history; the right column enlarges the terminal 18 reference cycles so that phase and waveform differences remain visible.

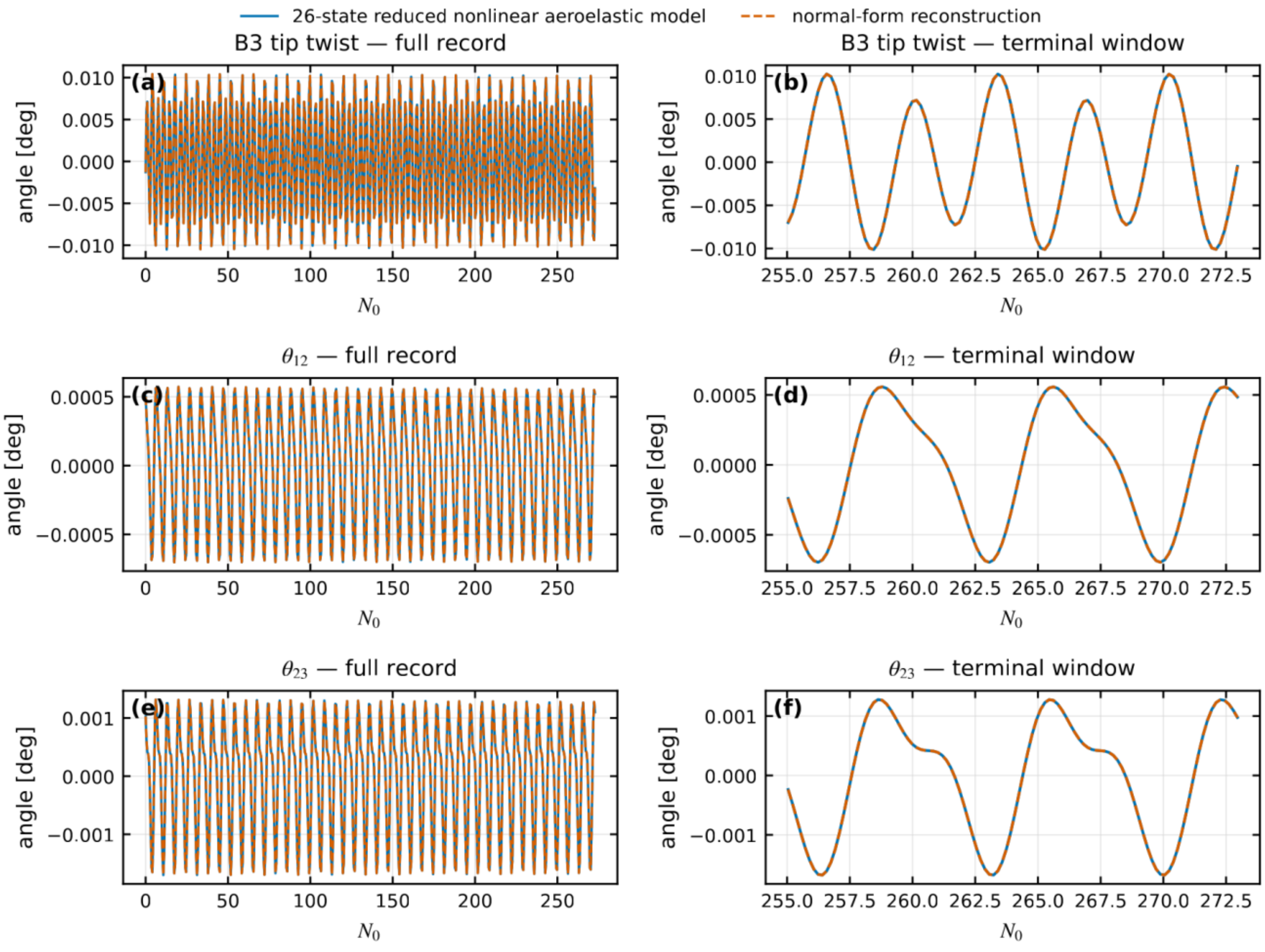

**Fig. 18. Four-dimensional normal-form reconstruction and 26-state model observations at the resonant point; columns show complete records and terminal windows.**

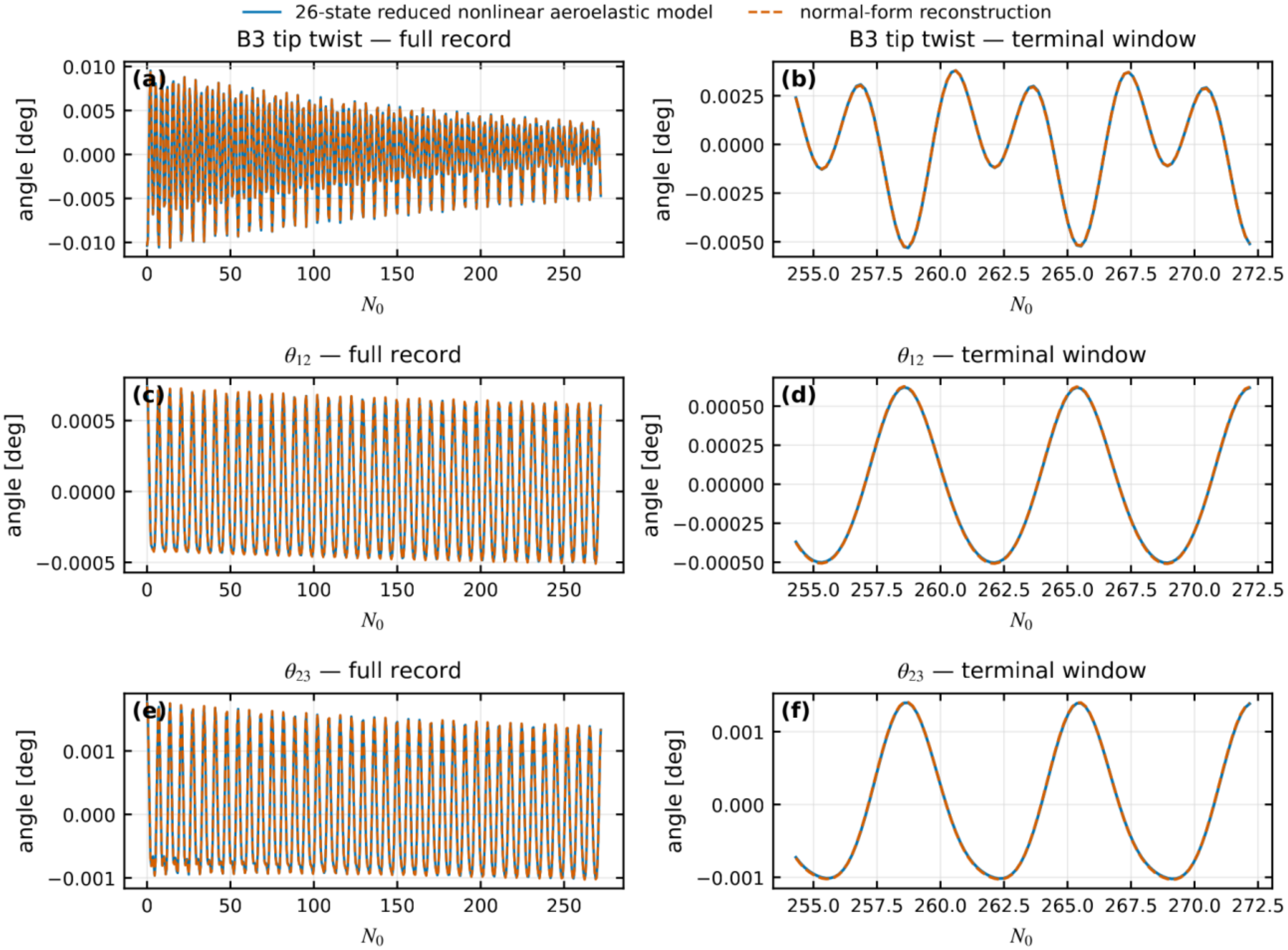

**Fig. 19. Four-dimensional normal-form reconstruction and 26-state model observations at the offset point; columns show complete records and terminal windows.**

For a sampled observation $y$, define

$$e_2 = 100\frac{\| y_{26} - y_{NF} \|_2}{\| y_{26} \|_2}\ \%, \qquad e_\infty = \max_k \left| y_{26,k} - y_{NF,k} \right| \tag{49}$$

Table 6 evaluates Eq. (49) over the complete records, not only the enlarged windows. At $P_0$, all normalized RMS errors are below 0.065%. The offset point produces the more demanding comparison: its largest value is 1.0427% for $\theta_{12}$, while the largest absolute angular difference is $8.698 \times 10^{-5}$ deg for B3 tip twist. The agreement is an internal reconstruction-consistency check for the selected observations and trajectories.

**Table 6. Full-record normal-form/26-state reconstruction differences.**

| Point | Observation | $e_2$ [%] | $e_\infty$ [deg] |
|---|---|---|---|
| $P_0$ | B3 tip twist | 0.06432 | $1.0168 \times 10^{-5}$ |
| $P_0$ | $\theta_{12}$ | 0.02014 | $2.6231 \times 10^{-7}$ |
| $P_0$ | $\theta_{23}$ | 0.02840 | $7.8372 \times 10^{-7}$ |
| $P_s$ | B3 tip twist | 0.75857 | $8.6982 \times 10^{-5}$ |

| Point | Observation | $e_2$ [%] | $e_\infty$ [deg] |
|---|---|---|---|
| $P_s$ | $\theta_{12}$ | 1.04268 | $7.6443 \times 10^{-6}$ |
| $P_s$ | $\theta_{23}$ | 0.91977 | $1.7335 \times 10^{-5}$ |

**B. Shooting solutions and Floquet stability**

A periodic solution of the four-dimensional cubic normal form satisfies

$$\widehat{\Phi}_T(z_0;\mu) - z_0 = 0, \qquad h(z_0) = 0 \tag{50}$$

where $h$ removes the autonomous time-shift degree of freedom [43]. At

$$(\epsilon_A, \phi, U) = (-0.2412,\ 12.61^\circ,\ 4.971\,\mathrm{m\,s^{-1}}) \tag{51}$$

Newton shooting of the cubic normal form yields the two closed trajectories in Fig. 20. Here high-frequency-dominant means $\max r_H/\max r_L > 10^5$ under the scale in Eq. (42). The mixed phase-locked orbit maintains nonzero L and H amplitudes and keeps the resonant phase combination within a 0.019-rad interval before closing over one period. Their normalized peak amplitudes (low, high) are $(2.07 \times 10^{-6}, 0.233193)$ and $(0.126278, 0.002904)$, respectively. Panel (d) enlarges the small H circle of the mixed orbit.

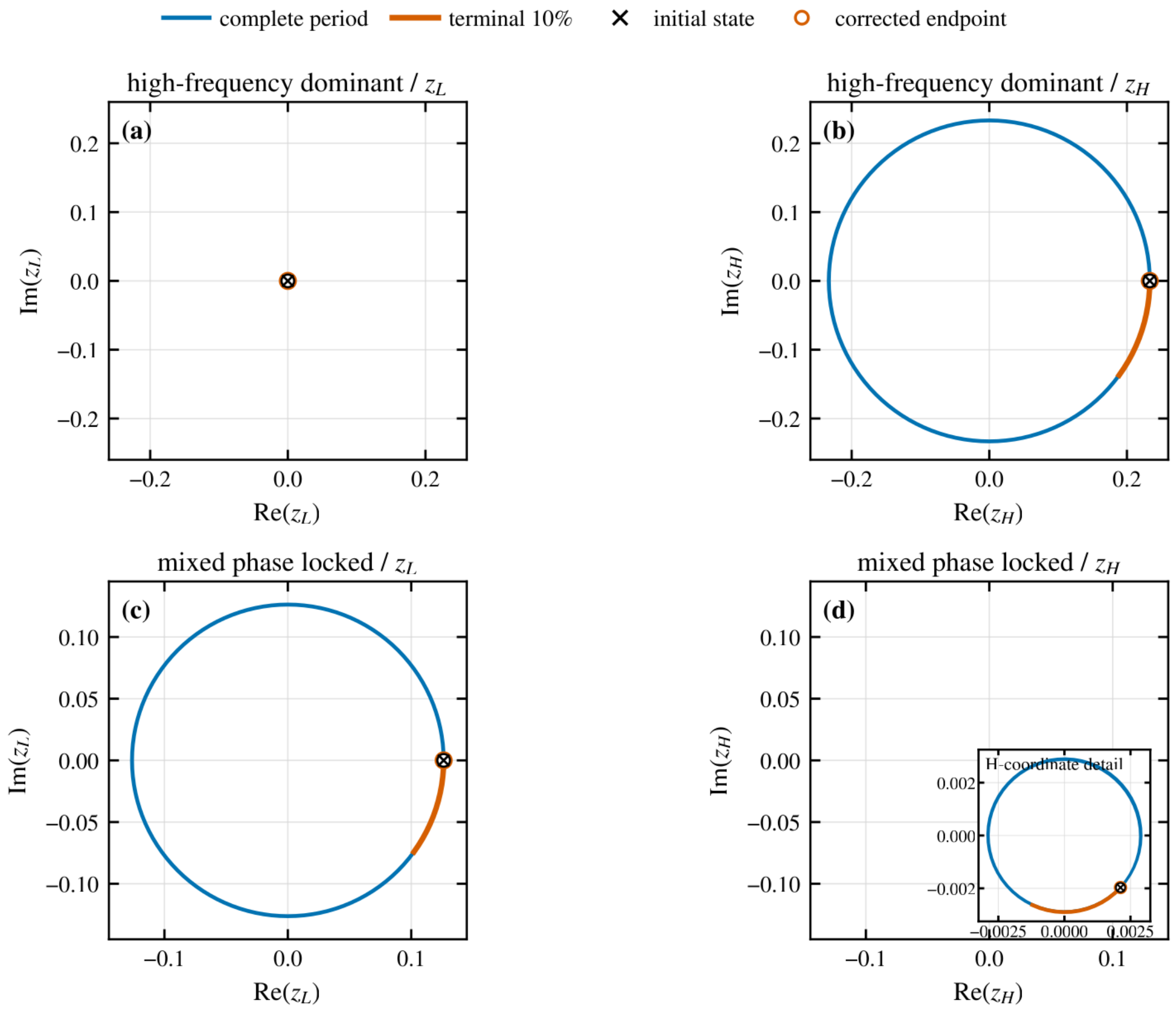

**Fig. 20. Periodic solutions admitted by the four-dimensional cubic normal form: a,b) high-frequency-dominant orbit; c,d) mixed phase-locked orbit with a high-frequency-coordinate inset.**

The monodromy matrix follows from

$$\dot{\Xi} = D_z \widehat{N}^{[3]} \left( z_p(t) \right) \Xi, \qquad \Xi(0) = I_4, \qquad \rho_F \in \operatorname{eig} \Xi(T) \tag{52}$$

One multiplier lies at unity because a time shift moves along the same autonomous orbit. Figure 21 retains the complete complex plane and unit circle for each solution. The largest nontrivial moduli are 0.999970 and 0.999852, and the remaining computed multipliers also lie inside the unit circle. Their proximity to unity implies only a small computed orbital-stability margin. Table 7 reports the Newton residual and independent relative endpoint closure.

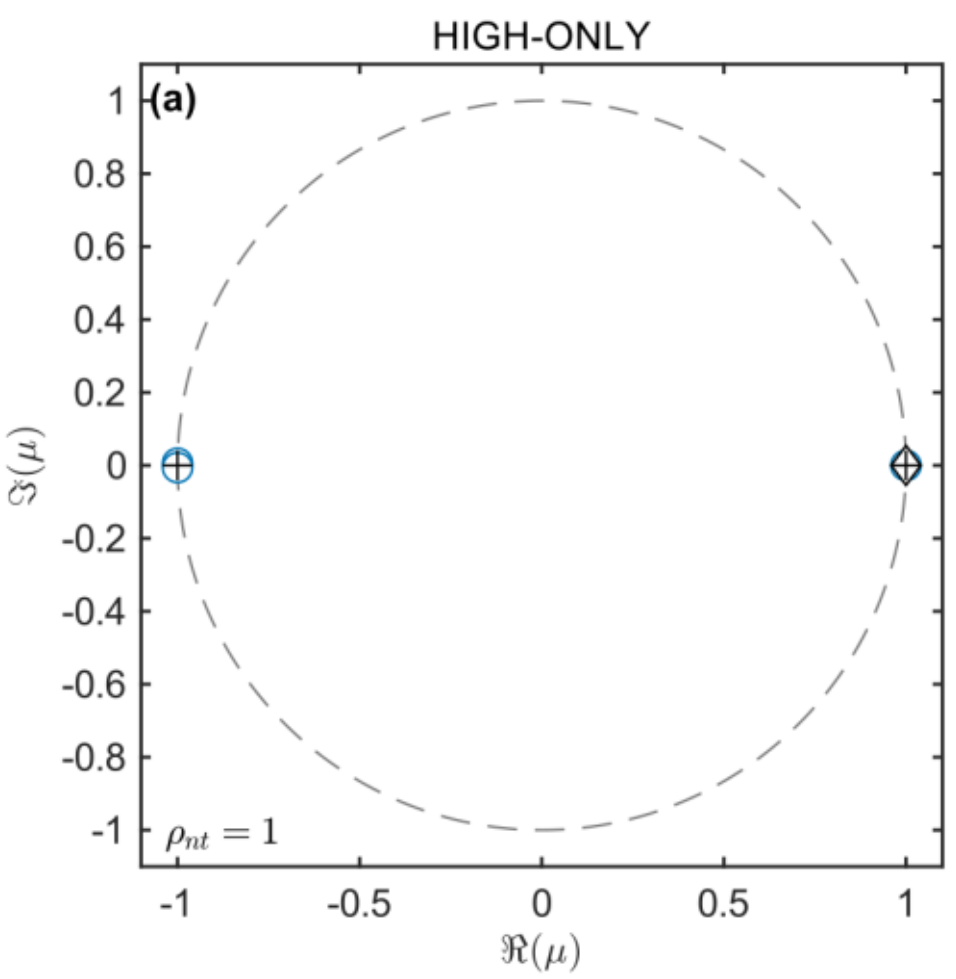

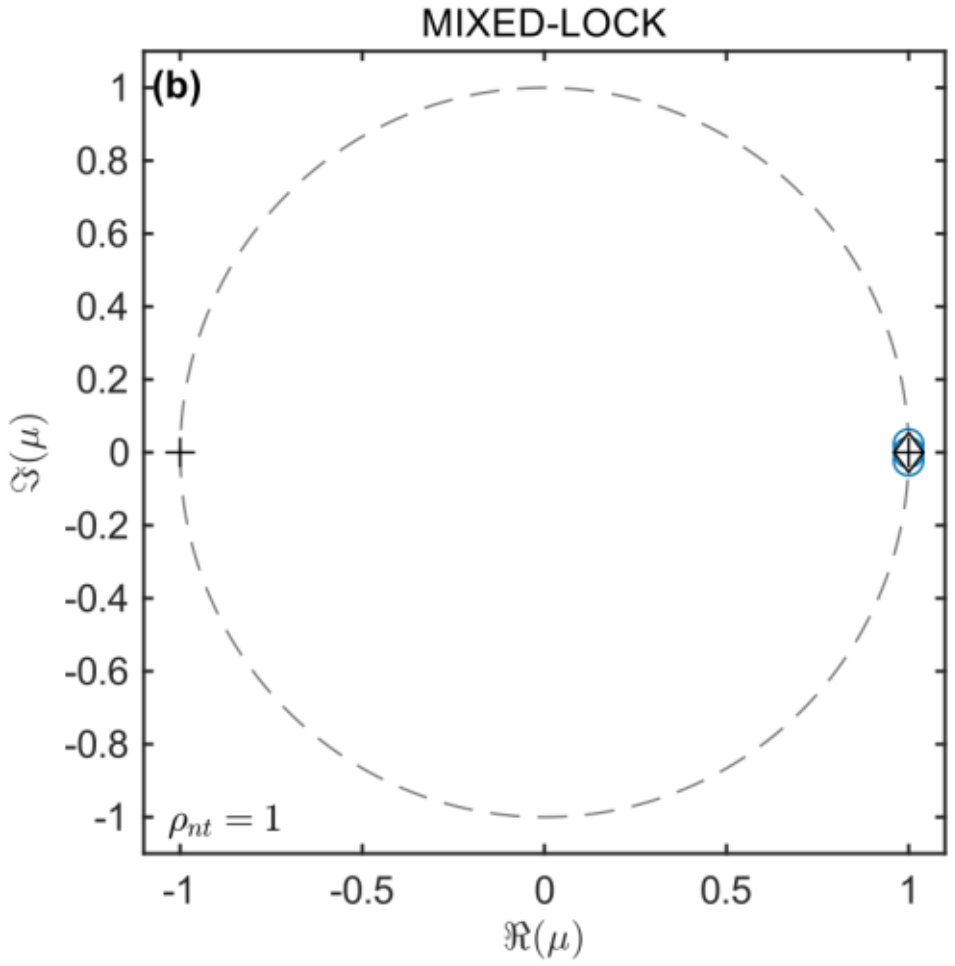

**Fig. 21. Floquet multipliers of the four-dimensional cubic normal form for a) the high-frequency-dominant and b) mixed phase-locked periodic solutions.**

**Table 7. Cubic-normal-form periodic-orbit closure and Floquet measures.**

| Quantity | High-frequency dominant | Mixed phase locked |
|---|---|---|
| Period T [s] | 0.758762601 | 1.51469681 |
| Shooting residual | $7.303 \times 10^{-9}$ | $9.989 \times 10^{-9}$ |
| Relative endpoint closure | $5.568 \times 10^{-8}$ | $1.852 \times 10^{-7}$ |
| Largest nontrivial \|ρF\| | 0.999970 | 0.999852 |

The period ratio is 1.99627. Together with the resonant phase closure and the spectra in Fig. 21, this value is consistent with the 1:2 organization found from the linear detuning and normal-form terms.

## VII. Local Cubic-Hinge Contribution

The local hinge correction is represented by the symmetric cubic torque

$$Q_{\theta_j}^{(3)} = -\beta\, \bar{k}_{\theta_j}^{(3)} \theta_j^3, \qquad j \in \{12,23\}, \qquad \beta \in \{0, 0.25, 0.762, 1.5\} \tag{53a}$$

The dimensional reference coefficient $\bar{k}_{\theta_j}^{(3)}$ has units $\mathrm{N\,m\,rad^{-3}}$, and $\beta$ is a dimensionless comparison multiplier. Projection of this cubic restoring-force tensor through the critical right eigenvectors and adjoint left eigenvectors changes the cubic normal-form coefficients, but not the quadratic 1:2 transfer terms. Figure 22(a,b) shows the six coefficient magnitudes and phases for the four values of $\beta$. Panels (c,d) show the associated changes of the modal amplitudes relative to $\beta = 0$. At $\beta = 1.5$, the largest coefficient-magnitude change is 0.763 ppm. Over the displayed trajectory, the maximum differences are $4.03 \times 10^{-9}$ in $r_L$ and $8.82 \times 10^{-10}$ in $r_H$.

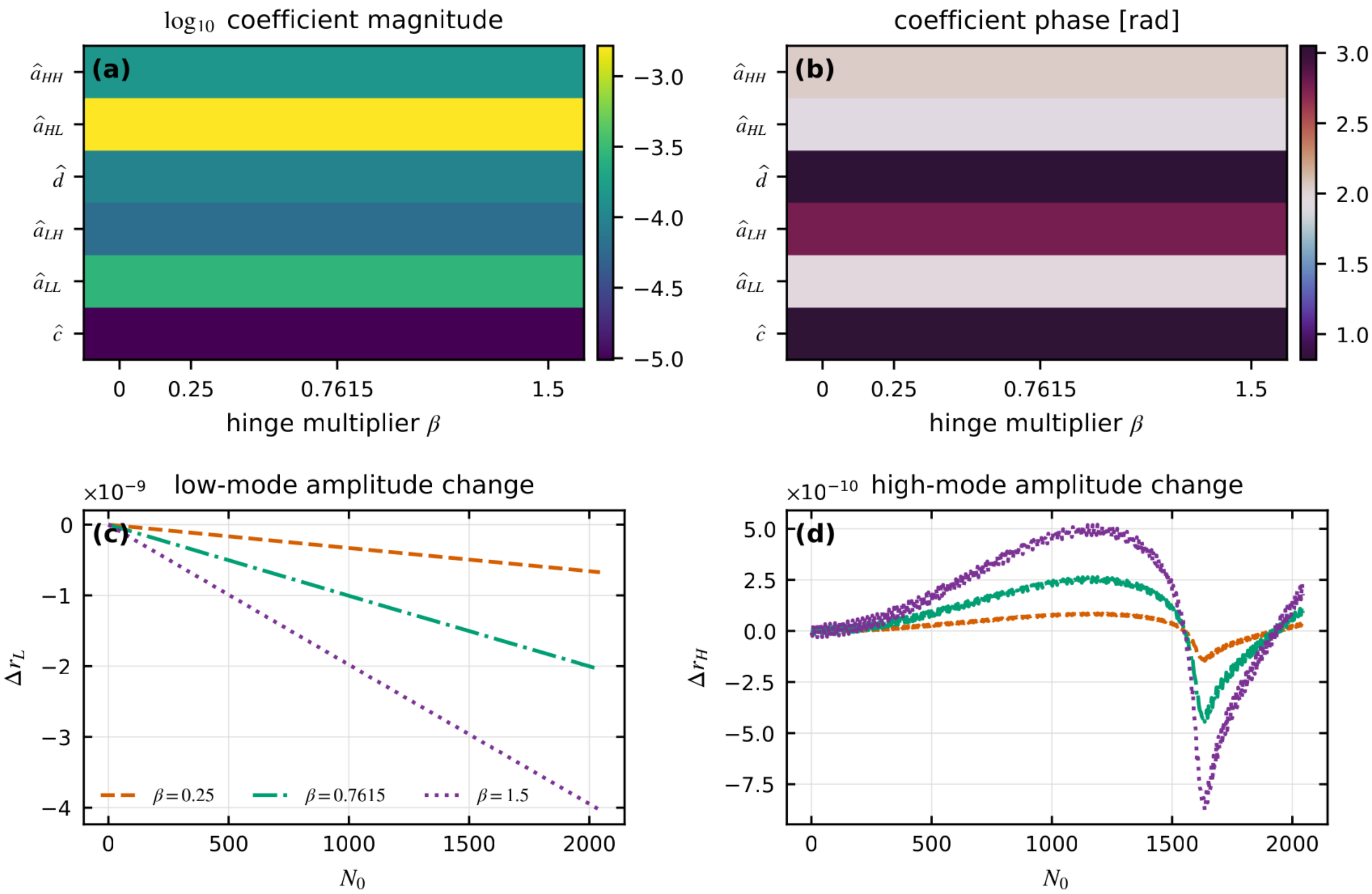

**Fig. 22. Cubic-hinge effects on a,b) normal-form coefficients and c,d) low- and high-frequency modal amplitudes.**

Within the adopted structural-restoring-force truncation and the tested neighborhood, the cubic hinge term provides a small correction to the final coefficients and trajectories. The retained distributed geometric-stiffness terms provide the dominant saturation contribution in this local model. The weak hinge channel indicated by the left–right participation and hinge-energy measures in Fig. 11 provides a structural interpretation.

Using the descriptor normalization in Eq. (22), let $\ell_{\theta_j}$ extract hinge coordinate $\theta_j$ from a state and let $b_{\theta_j}$ inject the associated work-conjugate generalized torque. A direct cubic hinge law enters a resonant coefficient through the same left–right channel that governs the linear hinge-stiffness sensitivity. Omitting the fixed symmetrization and coordinate-scaling factors already defined by the normal form, the two structures are

$$\gamma_{a;pqr}^{(\theta_j)} \propto \left(w_a^H b_{\theta_j}\right) \ell_{\theta_j}(v_p)\ell_{\theta_j}(v_q)\overline{\ell_{\theta_j}(v_r)}, \qquad \frac{\partial \lambda_a}{\partial k_{\theta_j}} \propto \left(w_a^H b_{\theta_j}\right) \ell_{\theta_j}(v_a) \tag{53b}$$

The left factor in Eq. (53b) measures how strongly the critical mode receives a hinge torque, whereas the coordinate-extraction factors measure hinge motion in the corresponding right modes. The direct cubic term carries three right-mode factors. Weak hinge motion therefore suppresses the local cubic contribution more strongly than the associated linear-stiffness channel under the declared amplitude and phase gauge [37, 38].

At the three representative crossings, Fig. 11 gives individual hinge-energy shares no greater than 1.3%, with the principal participation and elastic energy carried by B1 and B2. This weak hinge channel is consistent with the small coefficient and trajectory changes in Fig. 22. The participation and energy measures provide the structural interpretation, while the $\beta$ sweep supplies the quantitative local effect.

These results characterize the scheduling-point neighborhood and the amplitude scale declared in Eq. (42); they do not assign the same relative importance to hinge nonlinearity outside that local model.

## VIII. Conclusions

The component-attached formulation assembles geometrically exact component dynamics, joint transmission, aerodynamic-surface motion, pressure recovery, and virtual-work load pullback from explicit parent and material-coordinate maps. Its primitive-level descriptor separates structural, kinematic, dynamic-wake, and load-transfer actions while retaining exact speed scaling and algebraic assembly. This organization makes the source of a configuration action traceable and provides a route to blockwise residual, matrix, and eigenvalue derivatives.

The critical-frequency valley is a branch-competition phenomenon rather than a separate modal family. Fold angle changes relative surface–wake geometry and the projection of component motion onto local aerodynamic normals; the structural-axis offset changes attachment velocity and pressure-load lever arms. These paths redistribute left–right participation and strain energy primarily between $B_1$ and $B_2$. Consequently, the H, L, and H branches alternately form the lower neutral-speed envelope as fold angle increases, although the scalar flutter-speed curve remains smooth on the sampled grid.

Numerical continuation of the competing pairs localizes a near-1:2 double-Hopf point. Configuration-dependent critical eigenvectors change the projection of quadratic and cubic structural restoring-force tensors into the resonant normal form. Within that adopted truncation, the cubic normal form reproduces selected observations of the 26-state reduced model and admits high-frequency-dominant and mixed phase-locked periodic solutions. The reported Floquet spectra belong to this four-dimensional normal form. The cubic hinge term gives only a small local correction, consistent with the weak hinge participation and energy content.

Together, the structured parameter pathways, branch-resolved first-order information, and local resonant reduction distinguish both the branch that controls the flutter margin and the response classes admitted after the interacting modes approach neutrality. These distinctions can support configuration and stability-margin reasoning for morphing-wing design without reducing the analysis to a scalar boundary scan.

## Appendix A. Geometric Beam and Component-Graph Derivations

### A.1 Twist–wrench convention and Euler–Poincaré balance

Let $G = (R, r) \in SE(3)$ map material-section coordinates to the inertial frame. The body twist, body strain, and dual wrench use the order

$$\eta = \begin{bmatrix} v \\ \omega \end{bmatrix} = \left(G^{-1}G_{,t}\right)^{\vee}, \qquad \xi = \begin{bmatrix} \gamma \\ \kappa \end{bmatrix} = \left(G^{-1}G_{,s}\right)^{\vee}, \qquad \mathcal{B} = \begin{bmatrix} f \\ m \end{bmatrix}, \qquad \mathcal{B}^T\eta = f^T v + m^T\omega \tag{54}$$

For $a = [v^T, \omega^T]^T$, the adjoint algebra action in this ordering is

$$\mathrm{ad}_a = \begin{bmatrix} \hat{\omega} & \hat{v} \\ 0 & \hat{\omega} \end{bmatrix}, \qquad \mathrm{ad}_a^* = -\mathrm{ad}_a^T \tag{55}$$

A left-trivialized variation $\delta G = G\hat{\zeta}$ gives the exact identities

$$\delta\eta = \dot{\zeta} + \mathrm{ad}_\eta\zeta, \qquad \delta\xi = \zeta_{,s} + \mathrm{ad}_\xi\zeta \tag{56}$$

For kinetic density $\mathcal{T}(\eta)$, elastic density $\mathcal{V}(\xi)$, and distributed external wrench $f^{ext}$, define

$$\mu = \frac{\partial\mathcal{T}}{\partial\eta}, \qquad n = \frac{\partial\mathcal{V}}{\partial\xi}$$

The component variation is

$$\delta S = \int_{t_0}^{t_1}\int_0^L \left[\mu^T\left(\dot{\zeta} + \mathrm{ad}_\eta\zeta\right) - n^T\left(\zeta_{,s} + \mathrm{ad}_\xi\zeta\right) + (f^{ext})^T\zeta\right] ds\, dt \tag{57}$$

Integration by parts supplies the residual in Eq. (8), together with the material boundary one-form

$$-n^T\zeta\big|_{s=0}^{s=L} \tag{58}$$

This boundary term is the wrench port used by the component graph. The sign at a given endpoint follows from its outward material orientation; once that port orientation is fixed, all intercomponent transmission is a work-conjugate transpose operation.

### A.2 Internal velocity maps and serial assembly

For the factorization $H_i = B_i g_i E_i$, write

$$A_i = Ad_{g_i^{-1}}, \qquad \mathcal{E}_i = Ad_{E_i^{-1}}, \qquad D_i = dexp_{-\sigma_i}$$

Then the section velocity is

$$\eta_i = \mathcal{E}_i A_i V_{B_i} + D_i \dot{\sigma}_i \qquad (59)$$

After $\sigma_i = \Phi_i q_i$, any component port has the form introduced in Eq. (9), and its acceleration is

$$\dot{V}_p = Z_p \dot{V}_{B_i} + DZ_p[\dot{q}_i] V_{B_i} + Y_p \ddot{q}_i + DY_p[\dot{q}_i]\dot{q}_i \qquad (60)$$

Here $Z_p$ transports the inherited root motion and $Y_p$ injects retained elastic motion. For a revolute edge $e$ from a parent port to a child root,

$$V_{B_c} = H_e V_p + J_e \dot{\theta}_e, \qquad \dot{V}_{B_c} = H_e \dot{V}_p + DH_e\left[\dot{\theta}_e\right] V_p + J_e \ddot{\theta}_e + DJ_e\left[\dot{\theta}_e\right]\dot{\theta}_e \qquad (61)$$

The virtual-displacement relation is $\zeta_{B_c} = H_e \zeta_p + J_e \delta\theta_e$. Therefore, for a child-root one-form $\mathcal{B}_c$,

$$\mathcal{B}_p \mathrel{+}= H_e^T \mathcal{B}_c, \qquad R_{\theta_e} \mathrel{+}= J_e^T \mathcal{B}_c, \qquad \mathcal{B}_{B_i} \mathrel{+}= Z_p^T \mathcal{B}_p, \qquad R_{q_i} \mathrel{+}= Y_p^T \mathcal{B}_p \qquad (62)$$

For the Z-fold, the B3 contribution is first returned through joint 23 and accumulated at the B2 tip. That complete B2–B3 subtree is then returned through joint 12. This reverse topological order is the cotangent counterpart of the downstream pose and velocity recursion.

**A.3 Endpoint-preserving element geometry**

For element endpoints $G_a, G_b$, let

$$d = \log(G_a^{-1} G_b)^{\vee}, \qquad d_0 = d|_{reference}$$

The endpoint perturbation expansion is anchored at the finite reference value,

$$d = d_0 + d^{[1]}(\delta x_a, \delta x_b) + d^{[2]}(\delta x_a, \delta x_b) + \cdots + d^{[N_d]}(\delta x_a, \delta x_b) + \mathcal{O}(\| \delta x \|^{N_d+1}) \qquad (63)$$

An internal section at $s = s_a + \alpha L_e$ is reconstructed as

$$G(s) = G_a \exp\left(\widehat{\sigma(\alpha)}\right), \qquad \sigma(0) = 0, \qquad \sigma(1) = d \qquad (64)$$

The relative logarithm $d$ and its compatible material derivative supply the strain measure; $\sigma(\alpha)$ supplies internal geometry for inertia and attachment. Retaining these distinct roles preserves the exact finite reference strain while allowing a selected polynomial order for internal-station reconstruction.

Finally, if $H_i = B_i \bar{H}_i$ and an upstream coordinate enters only $B_i$, then

$$H_i^{-1} H_{i,s} = \bar{H}_i^{-1} \bar{H}_{i,s} \qquad (65)$$

Such a coordinate does not enter that component's strain energy, but it still enters material velocity, inertia, and external work through $B_i^{-1} \dot{B}_i$.

**Appendix B. Attached Surface, Normal Linearization, and Load Pullback**

Let a structural section have pose $G = (R, r)$, and let $\rho(s, \eta)$ be the attached material offset. The aerodynamic point is

$$X_A = r + R\rho, \qquad X_{A,u} = X_{A,s} s_{,u} + R\rho_{,\eta} \eta_{,u}, \qquad X_{A,v} = X_{A,s} s_{,v} + R\rho_{,\eta} \eta_{,v} \qquad (66)$$

With $u$ directed from leading to trailing edge and $v$ from component root to tip,

$$a = X_{A,u} \times X_{A,v}, \qquad n = \frac{a}{\parallel a \parallel}$$

The exact first variation of the surface normal is

$$\delta n = \frac{1}{\parallel a \parallel} (I - nn^T) \delta a, \qquad \delta a = \delta X_{A,u} \times X_{A,v} + X_{A,u} \times \delta X_{A,v} \qquad (67)$$

Linearizing the normal boundary velocity about a steady scheduling state gives

$$\delta u_n = (\delta n)^T (U_\infty - v_A) - n^T \delta v_A \qquad (68)$$

The first term includes the normal-rotation contribution to aerodynamic stiffness; the second supplies the station-motion and velocity channels. Collecting the displacement and velocity coefficients produces $N_q$ and $N_v$ in Eq. (13).

For a panel resultant $(f_A, m_A)$ at $X_A$,

$$\delta W_A = f_A^T \delta X_A + m_A^T \delta \vartheta_A \qquad (69)$$

At a structural station $X_S$, moment equivalence requires

$$f_S = f_A, \qquad m_S = m_A + (X_A - X_S) \times f_A \qquad (70)$$

Substitution of the structural virtual-displacement map gives Eq. (16). Thus the same geometric derivative that sends structural motion to an aerodynamic station returns force and moment through its transpose. The definition of $\epsilon_A$ in Eq. (17) follows from this physical lever arm.

## Appendix C. UVLM Elimination, Pressure, and Speed Scaling

At each control point, continuous no penetration is

$$n_p^T\left[u_{ind}(\Gamma_b,\Gamma_{w0},\Gamma_w)+U_p-V_p\right]=0 \qquad (71)$$

The discrete vortex-to-control-point map can be factored into oriented edge incidence, Biot–Savart evaluation, and normal projection. Partitioning the result yields Eqs. (14a)–(14b). Define

$$T_{shed}=-K_5^{-1}K_4, \qquad K_\Gamma=K_1+K_2T_{shed}$$

The bound and newly shed circulations are then

$$\Gamma_{w0}=T_{shed}\Gamma_b, \qquad \Gamma_b=K_\Gamma^{-1}(b_n-K_3\Gamma_w) \qquad (72a)$$

and the transported wake satisfies

$$\dot{\Gamma}_w=(K_6-K_7T_{shed}K_\Gamma^{-1}K_3)\Gamma_w+K_7T_{shed}K_\Gamma^{-1}b_n \qquad (72b)$$

A first-order row-transport relation,

$$\dot{\Gamma}_{m,j}=\frac{U}{\Delta x_w}\left(\Gamma_{m-1,j}-\Gamma_{m,j}\right) \qquad (73)$$

explains the linear speed scaling of the wake generator and input. Pressure recovery is defined through a signed pressure potential $\Pi_p$,

$$\Pi=C_\Gamma\Gamma_b+C_{\dot{\Gamma}}\dot{\Gamma}_b, \qquad \Delta p_p^{phys}=\rho_\infty\Pi_p, \qquad F_p=A_p\Delta p_p^{phys}n_p \qquad (74)$$

The circulation-rate part of Eq. (74) generates the pressure-rate channel after Eq. (72a) is differentiated. With fixed geometry and wake spacing measured in chord units, the convective operators scale as in Eq. (19), while the coefficient multiplying $\dot{\Gamma}_b$ is speed independent. This separation produces the $U^2$ aerodynamic-stiffness block and the $U$ aerodynamic-damping and wake blocks in Eq. (21).

For a structural normal-velocity input $u_s=N_qq+N_v\dot{q}$, the generalized aerodynamic transfer is

$$G_a(s)=C_w(sI-A_w)^{-1}B_w+D_w+s\dot{D}_w \qquad (75)$$

Wake reduction acts only on the strictly proper term. The common direct and pressure-rate terms explain why the total transfer error in Figs. 6 and 7 can be smaller than an error computed from the wake-only term.

## Appendix D. Static Condensation and Structural Polynomial Tensors

Partition a component displacement as $x_c = \Phi q + \Psi a$, where $q$ is retained and $a$ is condensed. Expand the component potential about the scheduling geometry,

$$\mathcal{U}(x_c) = \frac{1}{2}K_{ij}x_i x_j + \frac{1}{3!}K_{ijk}x_i x_j x_k + \frac{1}{4!}K_{ijkl}x_i x_j x_k x_l + \cdots \qquad (76)$$

The slave equilibrium is $g(q,a) = \Psi^T \nabla \mathcal{U}(\Phi q + \Psi a) = 0$. If $K_{aa} = \Psi^T K \Psi$ is nonsingular, the implicit-function theorem gives $a = h(q)$. Its quadratic and cubic coefficients satisfy

$$K_{aa} h_2(q,q) = -\frac{1}{2}\Psi^T K^{[3]}[\Phi q, \Phi q] \qquad (77a)$$

$$K_{aa} h_3(q,q,q) = -\Psi^T \left( K^{[3]}[\Phi q, \Psi h_2(q,q)] + \frac{1}{6}K^{[4]}[\Phi q, \Phi q, \Phi q] \right) \qquad (77b)$$

Substitution into $\bar{\mathcal{U}}(q) = \mathcal{U}\big(\Phi q + \Psi h(q)\big)$ produces a conservative polynomial restoring force $r(q) = \nabla \bar{\mathcal{U}}(q)$. A quadratic displacement manifold combined with the necessary geometric expansion supplies $\bar{\mathcal{U}}$ through quartic order and $r(q)$ through cubic order, which is the order retained in Eq. (40).

## Appendix E. Common Coordinates, Eigenvector Gauge, and Sensitivities

At each local parameter sample $a$, a balanced realization supplies wake directions $V_a, W_a$ and Hankel singular values $\Sigma_a$. Weighted collections

$$\mathcal{V} = \left[V_1 \Sigma_1^{1/2}, \dots, V_m \Sigma_m^{1/2}\right], \qquad \mathcal{W} = \left[W_1 \Sigma_1^{1/2}, \dots, W_m \Sigma_m^{1/2}\right] \qquad (78)$$

are biorthogonally factored to obtain the fixed $V_w, W_w$ used in Eq. (35). Only the dynamic-wake blocks $A_1, B_1, C_1$ are projected into this common coordinate. The blocks $D_1, \dot{D}_1, N_{q1}, N_v, M_s, C_s, K_s$ remain in their natural coordinates. This removes arbitrary wake-basis rotations before every member of $\mathcal{P}$ is represented by Eq. (36).

Differentiating the local chart gives, for any configuration parameter $\mu$,

$$\partial_\mu \hat{P} = \left(P_{10} + 2P_{20}\hat{\epsilon} + P_{11}\hat{\phi}\right)\partial_\mu \hat{\epsilon} + \left(P_{01} + P_{11}\hat{\epsilon} + 2P_{02}\hat{\phi}\right)\partial_\mu \hat{\phi}, \qquad P \in \mathcal{P}$$

Substitution into Eqs. (20)–(21) supplies $\partial_\mu E$ and $\partial_\mu L$ by the product rule. For speed, the same derivatives follow directly from the exact $U$ and $U^2$ factors. Equation (37) then completes the action from primitive blocks to a simple eigenvalue.

For the full coupled pencil, right and left eigenvectors are first normalized by Eq. (22). Given a preceding pair $(v_j^-, w_j^-)$ and a newly matched pair $(v_j^+, w_j^+)$, define

$$s_j = (w_j^-)^H E v_j^+, \qquad \tilde{v}_j^+ = e^{-i\arg s_j} v_j^+, \qquad \tilde{w}_j^+ = e^{-i\arg s_j} w_j^+ \tag{79}$$

Then $(w_j^-)^H E \tilde{v}_j^+ = |s_j|$ is positive real and $(\tilde{w}_j^+)^H E \tilde{v}_j^+ = 1$. Under a common rephasing of the matched right and left vectors, the block contractions in Eq. (33) are invariant. The complex center-coordinate coefficients nevertheless inherit the relative L/H phase convention, which is why Eq. (79) is also applied before the homological equations are assembled.

Equation (37) follows by differentiating $L v_j = \lambda_j E v_j$, premultiplying by $w_j^H$, and using the left eigenproblem. For the three double-Hopf conditions, a scaled Newton correction solves

$$J_F \Delta\mu = -F, \qquad F = [\beta_L \quad \beta_H \quad \Delta_{12}]^T, \qquad J_F = \frac{\partial F}{\partial(\epsilon_A, \phi, V^*)} \tag{80}$$

The singular values reported in Fig. 15 use both the raw $J_F$ and a version whose parameter columns are multiplied by the local box widths.

### Appendix F. Homological Equations, Reconstruction, and Floquet Test

Write Eq. (40) using symmetric multilinear maps $B$ and $C$, and expand

$$\mathcal{K}(z) = V_c z + K_2(z,z) + K_3(z,z,z) + \cdots, \qquad N(z) = \Lambda_c z + N_2(z,z) + N_3(z,z,z) + \cdots \tag{81}$$

At a monomial with multi-index $m$, the homological equation has the form

$$[L - (m \cdot \lambda)E] K_m = E V_c N_m - H_m \tag{82}$$

where $H_m$ is assembled from lower-order parameterization coefficients and the multilinear maps. For nonresonant monomials, $N_m$ is set to zero and Eq. (82) determines $K_m$. Near $\omega_H = 2\omega_L$, the monomials $\bar{z}_L z_H$ and $z_L^2$ are retained in $N_2$. The four cubic self- and cross-saturation monomials are retained in $N_3$, yielding Eqs. (44a)–(44b).

The physical reconstruction is the composition

$$z(0) \overset{N^{[3]}}{\rightarrow} z(\tau) \overset{\zeta = \rho_{work} z}{\rightarrow} \zeta(t) \overset{\mathcal{K}^{[3]}}{\rightarrow} x_{NF}(t) \overset{C}{\rightarrow} y_{NF}(t), \qquad \tau = \omega_0 t \tag{83}$$

This composition also initializes Eq. (40), as stated in Eq. (48). Because both integrations use the same mapped initial state, Table 6 isolates vector-field and truncation differences over the recorded trajectories.

For a corrected normal-form periodic orbit $z_p(t)$, Eq. (52) supplies the four-dimensional monodromy matrix. One multiplier is associated with $\dot{z}_p(0)$ and equals unity. The remaining multipliers determine orbital stability. When Floquet exponents are required, they are obtained from

$$\nu_j = \frac{1}{T}\log\rho_{F,j} \qquad (84)$$

with logarithm branches matched continuously to the quotient linearization.

## Appendix G. Supplementary Flutter Slices

Figures G1 and G2 give all six offset slices used to support Fig. 8. They use the identical baseline analysis grid, root definition, and frequency normalization. Separating them into two figures preserves the narrow frequency valley without reducing labels below the final-page reading size.

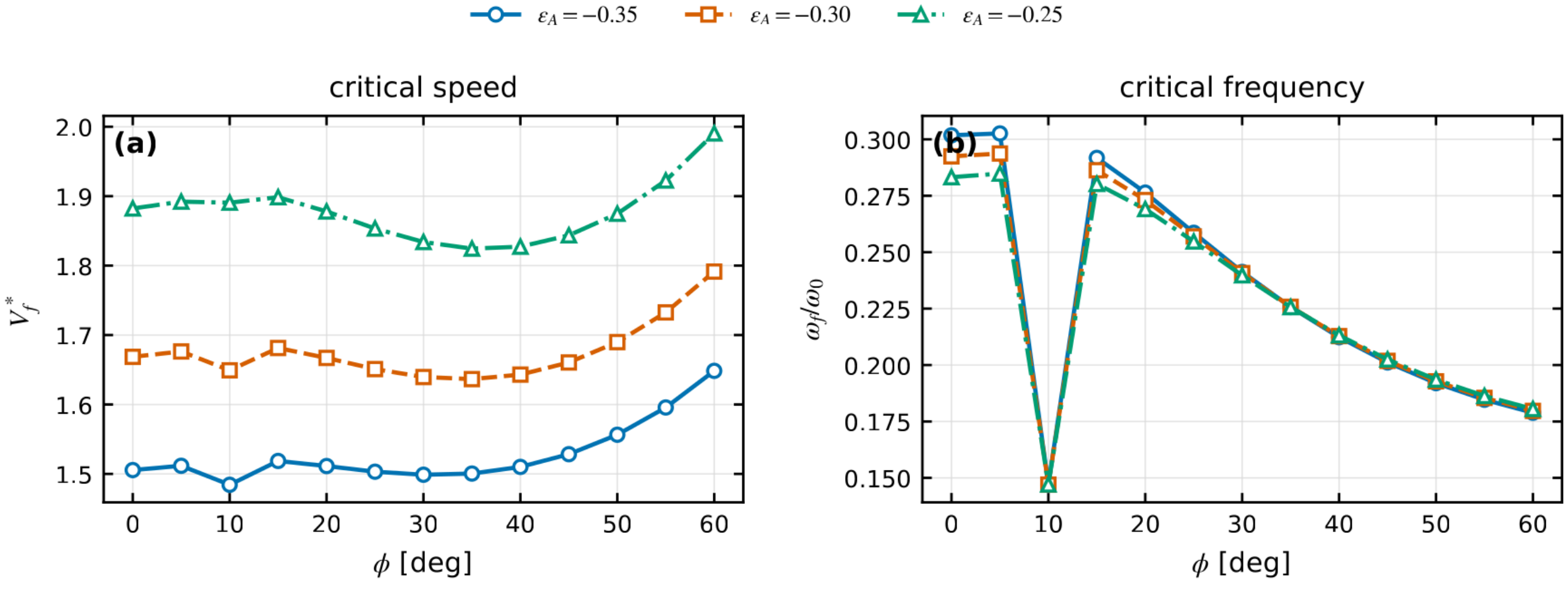

**Fig. G1. Flutter speed and critical-frequency slices for structural-axis offsets -0.35, -0.30, and -0.25.**

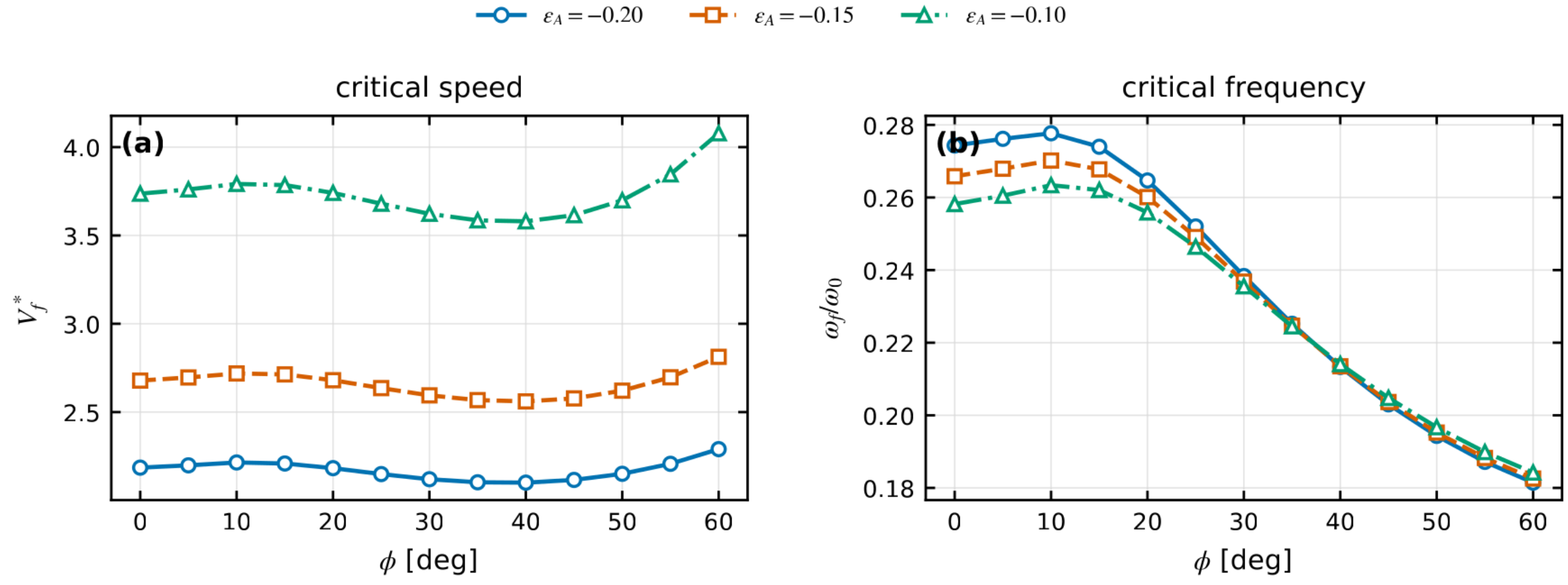

**Fig. G2. Flutter speed and critical-frequency slices for structural-axis offsets -0.20, -0.15, and -0.10.**

## Data Availability

Supporting data are available from the corresponding author upon reasonable request.

## Acknowledgments

Author contributions are as follows. Gelin Chen developed the methodology, defined the research problem and model configuration, derived the governing equations, and prepared the manuscript and figures. Haolin Yang contributed to manuscript writing and subsequent editing, reviewed and refined aeroelastic terminology, advised on figure and table formatting, and participated in discussions of the modeling formulation with the other authors. Chen Song contributed to discussions and advised on manuscript preparation. Chao Yang reviewed the manuscript for technical content and presentation.

During the preparation of this work, the authors used ChatGPT and Codex (OpenAI) to assist with language editing, code review, plotting, and manuscript workflow. The authors reviewed and edited all outputs and take full responsibility for the content of the publication.

## Funding

This work was supported by the National Natural Science Foundation of China (Grant No. 92471202).

## Conflict of Interest

The authors declare no competing interests relevant to this work.

## References

[1] Ajaj, R. M., Parancheerivilakkathil, M. S., Amoozgar, M., Friswell, M. I., and Cantwell, W. J., "Recent Developments in the Aeroelasticity of Morphing Aircraft," *Progress in Aerospace Sciences*, Vol. 120, 2021, 100682. https://doi.org/10.1016/j.paerosci.2020.100682

[2] Dai, Y., Li, J., Hu, Y., Zhang, J., Zhang, Y., Xi, Z., Zheng, Y., and Friswell, M. I., "Progress in Aerodynamics and Aeroelasticity of Morphing Aircraft," *Progress in Aerospace Sciences*, Vol. 163, 2026, 101206. https://doi.org/10.1016/j.paerosci.2026.101206

[3] Woods, B. K. S., Dayyani, I., and Friswell, M. I., "Fluid–Structure Interaction Analysis of the Fish Bone Active Camber Mechanism," *Journal of Aircraft*, Vol. 52, No. 1, 2015, pp. 307–319. https://doi.org/10.2514/1.C032725

[4] Zhang, J., Shaw, A. D., Wang, C., Gu, H., Amoozgar, M., Friswell, M. I., and Woods, B. K. S., "Aeroelastic Model and Analysis of an Active Camber Morphing Wing," *Aerospace Science and Technology*, Vol. 111, 2021, 106534. https://doi.org/10.1016/j.ast.2021.106534

[5] Tang, D., and Dowell, E. H., "Theoretical and Experimental Aeroelastic Study for Folding Wing Structures," *Journal of Aircraft*, Vol. 45, No. 4, 2008, pp. 1136–1147. https://doi.org/10.2514/1.32754

[6] Liska, S., and Dowell, E. H., "Continuum Aeroelastic Model for a Folding-Wing Configuration," *AIAA Journal*, Vol. 47, No. 10, 2009, pp. 2350–2358. https://doi.org/10.2514/1.40475

[7] Attar, P. J., Tang, D., and Dowell, E. H., "Nonlinear Aeroelastic Study for Folding Wing Structures," *AIAA Journal*, Vol. 48, No. 10, 2010, pp. 2187–2195. https://doi.org/10.2514/1.44868

[8] Wang, I., Gibbs, S. C., and Dowell, E. H., "Aeroelastic Model of Multisegmented Folding Wings: Theory and Experiment," *Journal of Aircraft*, Vol. 49, No. 3, 2012, pp. 911–921. https://doi.org/10.2514/1.C031589

[9] Verstraete, M., Roccia, B. A., Mook, D. T., and Preidikman, S., "A Co-Simulation Methodology to Simulate the Nonlinear Aeroelastic Behavior of a Folding-Wing Concept in Different Flight Configurations," *Nonlinear Dynamics*, Vol. 98, 2019, pp. 907–927. https://doi.org/10.1007/s11071-019-05234-9

[10] Xie, C., Chen, Z., and An, C., "Aeroelastic Response of a Z-Shaped Folding Wing During the Morphing Process," *AIAA Journal*, Vol. 60, No. 5, 2022, pp. 3166–3179. https://doi.org/10.2514/1.J061138

[11] Yang, H., Song, C., Yang, C., Chen, G., and Dang, Q., "Flutter Analysis for Folding-Wing Aircraft Considering Rigid–Flexible Coupling," *AIAA Journal*, published online 24 August 2026. https://doi.org/10.2514/1.J066663

[12] Zhao, Y. H., and Hu, H. Y., "Parameterized Aeroelastic Modeling and Flutter Analysis for a Folding Wing," *Journal of Sound and Vibration*, Vol. 331, No. 2, 2012, pp. 308–324. https://doi.org/10.1016/j.jsv.2011.08.028

[13] Hu, W., Yang, Z., and Gu, Y., "Aeroelastic Study for Folding Wing During the Morphing Process," *Journal of Sound and Vibration*, Vol. 365, 2016, pp. 216–229. https://doi.org/10.1016/j.jsv.2015.11.043

[14] Hu, W., Yang, Z., Gu, Y., and Wang, X., "The Nonlinear Aeroelastic Characteristics of a Folding Wing with Cubic Stiffness," *Journal of Sound and Vibration*, Vol. 400, 2017, pp. 22–39. https://doi.org/10.1016/j.jsv.2017.04.002

[15] Huang, R., and Zhou, X., "Parameterized Fictitious Mode of Morphing Wing with Bilinear Hinge Stiffness," *AIAA Journal*, Vol. 59, No. 7, 2021, pp. 2641–2656. https://doi.org/10.2514/1.J059347

[16] Guo, X., Wang, S., Qu, Y., and Cao, D., "Nonlinear Dynamics of Z-Shaped Morphing Wings in Subsonic Flow," *Aerospace Science and Technology*, Vol. 119, 2021, 107145. https://doi.org/10.1016/j.ast.2021.107145

[17] Guo, X., Zhang, Y., Zhang, W., Sun, L., and Chen, S., "Nonlinear Dynamics of Z-Shaped Folding Wings with 1:1 Inner Resonance," *International Journal of Bifurcation and Chaos*, Vol. 27, No. 8, 2017, 1750124. https://doi.org/10.1142/S0218127417501243

[18] Murua, J., Palacios, R., and Graham, J. M. R., "Applications of the Unsteady Vortex-Lattice Method in Aircraft Aeroelasticity and Flight Dynamics," *Progress in Aerospace Sciences*, Vol. 55, 2012, pp. 46–72. https://doi.org/10.1016/j.paerosci.2012.06.001

[19] Maraniello, S., and Palacios, R., "State-Space Realizations and Internal Balancing in Potential-Flow Aerodynamics with Arbitrary Kinematics," *AIAA Journal*, Vol. 57, No. 6, 2019, pp. 2308–2321. https://doi.org/10.2514/1.J058153

[20] Fonzi, N., Brunton, S. L., and Fasel, U., "Data-Driven Nonlinear Aeroelastic Models of Morphing Wings for Control," *Proceedings of the Royal Society A*, Vol. 476, No. 2239, 2020, 20200079. https://doi.org/10.1098/rspa.2020.0079

[21] Maraniello, S., and Palacios, R., "Parametric Reduced-Order Modeling of the Unsteady Vortex-Lattice Method," *AIAA Journal*, Vol. 58, No. 5, 2020, pp. 2206–2220. https://doi.org/10.2514/1.J058894

[22] Stewart, G. W., and Sun, J.-G., *Matrix Perturbation Theory*, Academic Press, Boston, MA, 1990.

[23] Gattulli, V., Di Fabio, F., and Luongo, A., “One to One Resonant Double Hopf Bifurcation in Aeroelastic Oscillators with Tuned Mass Dampers,” *Journal of Sound and Vibration*, Vol. 262, No. 2, 2003, pp. 201–217. https://doi.org/10.1016/S0022-460X(02)01135-5

[24] Nayfeh, A. H., Ghommem, M., and Hajj, M. R., “Normal Form Representation of the Aeroelastic Response of the Goland Wing,” *Nonlinear Dynamics*, Vol. 67, 2012, pp. 1847–1861. https://doi.org/10.1007/s11071-011-0111-6

[25] Wu, X., Li, G., and Yue, Y., “Effect of Hopf-Hopf Bifurcation on the Post-Flutter Behavior of a Three-Degree-of-Freedom Airfoil,” *Aerospace Science and Technology*, Vol. 154, 2024, 109525. https://doi.org/10.1016/j.ast.2024.109525

[26] Luongo, A., Paolone, A., and Di Egidio, A., “Multiple Timescales Analysis for 1:2 and 1:3 Resonant Hopf Bifurcations,” *Nonlinear Dynamics*, Vol. 34, 2003, pp. 269–291. https://doi.org/10.1023/B:NODY.0000013508.50435.39

[27] Simo, J. C., “A Finite Strain Beam Formulation. The Three-Dimensional Dynamic Problem. Part I,” *Computer Methods in Applied Mechanics and Engineering*, Vol. 49, No. 1, 1985, pp. 55–70. https://doi.org/10.1016/0045-7825(85)90050-7

[28] Hodges, D. H., “A Mixed Variational Formulation Based on Exact Intrinsic Equations for Dynamics of Moving Beams,” *International Journal of Solids and Structures*, Vol. 26, No. 11, 1990, pp. 1253–1273. https://doi.org/10.1016/0020-7683(90)90060-9

[29] Hodges, D. H., “Geometrically Exact, Intrinsic Theory for Dynamics of Curved and Twisted Anisotropic Beams,” *AIAA Journal*, Vol. 41, No. 6, 2003, pp. 1131–1137. https://doi.org/10.2514/2.2054

[30] Bauchau, O. A., *Flexible Multibody Dynamics*, Springer, Dordrecht, The Netherlands, 2011. https://doi.org/10.1007/978-94-007-0335-3

[31] Holm, D. D., Marsden, J. E., and Ratiu, T. S., “The Euler–Poincaré Equations and Semidirect Products with Applications to Continuum Theories,” *Advances in Mathematics*, Vol. 137, No. 1, 1998, pp. 1–81. https://doi.org/10.1006/aima.1998.1721

[32] Hente, C., Roccia, B. A., Rolfes, R., and Gebhardt, C. G., “Analytical Linearization of Aerodynamic Loads in Unsteady Vortex-Lattice Method for Nonlinear Aeroelastic Applications,” *AIAA Journal*, Vol. 62, No. 10, 2024, pp. 3857–3880. https://doi.org/10.2514/1.J063693

[33] Katz, J., and Plotkin, A., *Low-Speed Aerodynamics*, 2nd ed., Cambridge University Press, Cambridge, UK, 2001. https://doi.org/10.1017/CBO9780511810329

[34] Palacios, R., Murua, J., and Cook, R., "Structural and Aerodynamic Models in Nonlinear Flight Dynamics of Very Flexible Aircraft," *AIAA Journal*, Vol. 48, No. 11, 2010, pp. 2648–2659. https://doi.org/10.2514/1.J050513

[35] Hesse, H., Palacios, R., and Murua, J., "Consistent Structural Linearization in Flexible Aircraft Dynamics with Large Rigid-Body Motion," *AIAA Journal*, Vol. 52, No. 3, 2014, pp. 528–538. https://doi.org/10.2514/1.J052316

[36] Hashlamoun, W. A., Hassouneh, M. A., and Abed, E. H., "New Results on Modal Participation Factors: Revealing a Previously Unknown Dichotomy," *IEEE Transactions on Automatic Control*, Vol. 54, No. 7, 2009, pp. 1439–1449. https://doi.org/10.1109/TAC.2009.2019796

[37] Guckenheimer, J., and Holmes, P., *Nonlinear Oscillations, Dynamical Systems, and Bifurcations of Vector Fields*, Springer, New York, 1983.

[38] Kuznetsov, Y. A., *Elements of Applied Bifurcation Theory*, 3rd ed., Springer, New York, 2004. https://doi.org/10.1007/978-1-4757-3978-7

[39] Nayfeh, A. H., and Mook, D. T., *Nonlinear Oscillations*, Wiley, New York, 1995.

[40] Lee, B. H. K., Price, S. J., and Wong, Y. S., "Nonlinear Aeroelastic Analysis of Airfoils: Bifurcation and Chaos," *Progress in Aerospace Sciences*, Vol. 35, No. 3, 1999, pp. 205–334. https://doi.org/10.1016/S0376-0421(98)00015-3

[41] Haller, G., and Ponsioen, S., "Nonlinear Normal Modes and Spectral Submanifolds: Existence, Uniqueness and Use in Model Reduction," *Nonlinear Dynamics*, Vol. 86, No. 3, 2016, pp. 1493–1534. https://doi.org/10.1007/s11071-016-2974-z

[42] Jain, S., and Haller, G., "How to Compute Invariant Manifolds and Their Reduced Dynamics in High-Dimensional Finite-Element Models," *Nonlinear Dynamics*, Vol. 107, 2022, pp. 1417–1450. https://doi.org/10.1007/s11071-021-06957-4

[43] Peeters, M., Viguié, R., Sérandour, G., Kerschen, G., and Golinval, J.-C., "Nonlinear Normal Modes, Part II: Toward a Practical Computation Using Numerical Continuation Techniques," *Mechanical Systems and Signal Processing*, Vol. 23, No. 1, 2009, pp. 195–216. https://doi.org/10.1016/j.ymssp.2008.04.003